\documentclass[fleqn,usenatbib]{mnras}

\usepackage{graphicx}	
\usepackage{amsmath}	
\usepackage{nccmath}
\usepackage{amssymb}	
\usepackage{caption}
\usepackage{subcaption}

\usepackage{smartdiagram}
\usesmartdiagramlibrary{additions}
\usepackage{multirow}
\usepackage{multicol}

\title[Synchro-curvature $\gamma$-ray light curves and spectra]
{
Synchro-curvature description of $\gamma$-ray light curves and spectra of pulsars: global properties
}

\author[\'I\~niguez-Pascual, Torres \& Vigan\`o]{
Daniel \'I\~niguez-Pascual$^{1,2}$\thanks{E-mail: iniguez@ice.csic.es}, Diego F. Torres$^{1,2,3}$\thanks{E-mail: dtorres@ice.csic.es},
Daniele Vigan\`o$^{1,2,4}$\thanks{E-mail: vigano@ice.csic.es}
\\
$^{1}$Institute of Space Sciences (ICE, CSIC), Campus UAB, Carrer de Can Magrans s/n, 08193 Barcelona, Spain\\
$^{2}$Institut d’Estudis Espacials de Catalunya (IEEC), 08034 Barcelona, Spain\\
$^{3}$Institució Catalana de Recerca i Estudis Avançats (ICREA), E-08010 Barcelona, Spain \\
$^{4}$Institute of Applied Computing \& Community Code (IAC3), University of the Balearic Islands, Palma, 07122, Spain
}
\date{}

\pubyear{2023}

\begin{document}
\label{firstpage}
\pagerange{\pageref{firstpage}--\pageref{lastpage}}
\maketitle

\begin{abstract}
This work presents a methodological approach to generate realistic $\gamma$-ray light curves of pulsars, resembling reasonably well the observational ones observed by the {\it Fermi}-Large Area Telescope instrument, fitting at the same time their high-energy spectra. 
The theoretical light curves are obtained from a spectral and geometrical model of the synchro-curvature emission. 
Despite our model relies on a few effective physical parameters, the synthetic light curves present the same main features observed in the observational $\gamma$-ray light curve zoo, such as the different shapes, variety in the number of peaks, and a diversity of peak widths.
The morphological features of the light curves allows us to statistically compare the observed properties. 
In particular, we find that the proportion on the number of peaks found in our synthetic light curves is in agreement with the observational one provided by the third Fermi-LAT pulsar catalog. 
We also found that the detection probability due to beaming is much higher for orthogonal rotators (approaching 100\%) than for small inclination angles (less than 20\%).
The small variation on the synthetic skymaps generated for different pulsars indicates that the geometry dominates over timing and spectral properties in shaping the gamma-ray light curves.
This means that geometrical parameters like the inclination angle can be in principle constrained by gamma-ray data alone independently on the specific properties of a pulsar.
At the same time, we find that $\gamma$-ray spectra seen by different observers can slightly differ, opening the door to 
constraining the viewing angle of a particular pulsar.

\end{abstract}

\begin{keywords}  
pulsars: general -- gamma-rays: stars -- acceleration of particles -- radiation mechanisms: non-thermal
\end{keywords}

\section{Introduction}
\label{introduction}

Modeling $\gamma$-ray light curves of pulsars is a very powerful tool to understand the geometry of the accelerating regions of the magnetosphere where their high-energy radiation is produced.
However, the exact locations and shape of such regions, where the force-free condition breaks and particle acceleration occurs, are not yet well determined.

In the past, many models proposed that these accelerating regions were located in so-called gaps, where a charge density smaller than the Goldreich-Julian one \citep{pulsar_electrodynamics} is not enough to screen the parallel electric field.
Relevant gap models included the Polar Cap  \citep{Sturrock71, Daugherty96}, the Slot Gap \citep{arons83, muslimov04}, and the Outer Gap \citep{cheng86, romani95}.
All of them place such gaps on top of the retarded vacuum solution from \cite{deutsch55}, so they differ mainly on where the gaps are located.
The advent of computationally advanced simulations shed new light on this topic over the last two decades.
Force-free electrodynamics simulations of the neutron star magnetosphere noted the Y-point and the equatorial current sheet as the place for particle acceleration \citep{Contopoulos99, Spitkovsky06}, presumably through continuous magnetic reconnection of open magnetic field lines \citep{Lyubarskii96}.
Advanced particle-in-cell (PIC) simulations, in which the particles populating the magnetosphere are self-consistently simulated together with the electromagnetic fields, found very similar results (see \citealt{Cerutti19} for a review).

Several attempts have been made to extract information on the high-energy emission topology from the $\gamma$-ray light curves, either assuming some or all of the gap models \citep{Watters09, Pierbattista15}, or considering numerical magnetospheric solutions \citep{Bai10b, Kalapotharakos14, Cao19}.
However, most of the models assume a uniform high-energy photon emission from the whole region and do not study its energy-dependence, being few those that compute the radiation emitted in the region as a function of position and energy \citep{Cerutti16}.
Even in these cases, coupling a PIC model with radiation has proven to be numerically challenging both for the scale ranges and the computational time involved.

In this paper, we significantly extend a geometrical model presented in \cite{light_curves_2019}, which introduced a novel way to geometrically represent these magnetospheric emission regions of pulsars. 
The approach consisted in using Frenet-Serret equations to geometrically describe the trajectory of the particles traveling the emission region.
Relying on the foundations established in that paper, we here make the topology of the emission region more realistic, consider torsion,
and implement the full synchro-curvature radiation mechanism with which the traveling particles emit high-energy radiation.
With the extended model presented in this work, we show it is possible to build emission maps (or skymaps), from which light curves can be obtained for a variety of geometrical and spectral parameters.
We couple these results to our earlier-introduced spectral model (see e.g., \cite{diego_solo,systematic_2019,sc_emitting_regions_2022})
so as to concurrently describe both light curves and spectra of pulsars.

This paper is organized as follows. 
Section \ref{geometrical_model} presents the radiative and geometrical model that leads to synthetic emission maps and light curves. 
In Section \ref{SMs_LCs} we show several examples of them, for some selected pulsars for which high-quality gamma-ray spectra constrain the input parameters of our radiative model. 
We present, in Section \ref{light_curves_features}, a way to qualitatively perform an analysis of the synthetic sample of light curves generated and show results for selected pulsars. 
Finally, in Section \ref{conclusions} we draw our main conclusions and describe our future research direction.

\section{Radiative and geometrical modeling}\label{geometrical_model}

\subsection{Trajectories}

Our model starts from a simple description of charged particle dynamics in an inertial lab frame. 
In this paper, we assume that the emission region is just outside the light cylinder, close to the Y-point where the force-free condition breaks and particles can be accelerated. 
In general, particles locally slide along the magnetic field lines and gyrate around them, with a Larmor radius much smaller than the magnetic lengthscale. Due to the drift, the trajectories drawn by a given charged particles deviate from the static magnetic field line \citep{Bai10b}.
Such deviation is even more substantial beyond the light cylinder, since the plasma cannot co-rotate with the neutron stars and the magnetic field lines are twisted.
Hereafter we will use $\lambda$ as the position  along the gyration-averaged trajectory (i.e., neglecting its gyrating component), or {\em along the rotating field lines}, e.g. along the magnetic field line taking into account the drift given by the strong rotation velocity of the plasma in which the field lines are frozen.
When we talk about the emission region, we refer to the zone where the trajectories of the particles are located, always from the point of view of an observer in the lab frame.

Having this in mind the dynamics of charged accelerated particles in non-force-free regions of the magnetosphere results  from the combination of the parallel electric field magnitude $E_{||}$, and the synchro-curvature losses coming from the combined gyration (around the lines) and parallel motion (along them). Such losses are, in turn, defined by the particle Lorentz factor $\Gamma$, the pitch angle $\alpha$ (angle between the rotating magnetic field line direction and the particle's direction of motion), and by the local values of the magnetic field, $B$, and the curvature radius of the gyration-averaged trajectory, $r_c$ \citep{compact_formulae}.
The latter two parameters are assumed to depend on the pulsars's spin period $P$ and time derivative $\Dot{P}$ and, in our model, are parametrized for simplicity as functions of the distance along $\lambda$, as $r_c=R_{lc} \, (x_{in}R_{lc} + \lambda/R_{lc})^{\eta}$ and 
$B = B_s (R_s /(x_{in}R_{lc} + \lambda))^{b}$, 
where $\eta$ is fixed to $0.5$, $b$ is the magnetic gradient 
(see \citealt{outer_gap_model_paper_2} for details), $x_{in}$ is the injection point distance (see discussion below), $R_{lc}$ is the light cylinder radius, $B_s$ is the timing-estimated magnitude of the poloidal dipolar component of the magnetic field at the polar surface,  and $R_s$ is the radius of the neutron star.

The electrical acceleration and the radiative losses are balanced by the equations of motion of a charged particle, which follow the variation of its relativistic momentum along the trajectory of the particle, see \cite{compact_formulae} and \cite{Hirotani99}.
The numerical solution of these equations gives the evolution along the trajectory of $\alpha$ and $\Gamma$. This, in turn, allows to calculate the spectral distribution of the emitted photons at each point of a particle trajectory. Note in the simplified particle dynamics we solve, that we neglect the drift term, which might have effects on the pitch angle (which in our equations rapidly approaches negligible values).

The motion of the particles is initially characterized by a non-negliglible gyration component, with relatively moderate Lorentz factors producing a peak of the emission in X-rays \citep{diego_solo}. 
While the particle parallel momentum increases due to the acceleration, the perpendicular momentum is rapidly radiated away and the pitch angle exponentially decreases, so that the particles effectively slide along the rotating magnetic field line and can reach Lorentz factors larger than $10^7$, with the radiation peaking at $\gamma$-rays. 

Complementary to this spectral model,
\cite{light_curves_2019} presented a new approach to geometrically represent these emission regions with a few effective parameters and obtain light curves for different observers.
It uses the Frenet-Serret differential geometry formulae, which allow one to geometrically describe the trajectory of a particle moving in a three-dimensional space.
At each $\lambda$, the gyration-average trajectory defines three directions: tangent $\hat{t}$, normal $\hat{n}$ (which points towards the center of curvature),
and binormal $\hat{b} = \hat{t} \times \hat{n}$. 
Having the curvature radius $r_c$ and the torsion $\tau$ as functions of the position $\lambda$, the evolution of these vectors is determined by the Frenet-Serret equations:
\begin{eqnarray}
\frac{d\hat{t}}{d\lambda} &=& \frac{1}{r_c} \hat{n}~, \nonumber \\
\frac{d\hat{n}}{d\lambda} &=& - \frac{1}{r_c} \hat{t} +\tau \hat{b}~, \nonumber \\
\frac{d\hat{b}}{d\lambda} &=& - \tau \hat{n}~.
\end{eqnarray}
We solve these equations with a fourth-order Runge-Kutta method sampling thousands of trajectory points, with a spatial griding which increase as the particles get faster, in order to capture better the initial, slower, synchrotron-dominated part. 

\cite{light_curves_2019} assumed zero torsion, which is valid as long as the twist of the magnetic field lines is small compared to its curvature. 
Here, we shall get rid of such assumption and allow curvature and torsion being of the same order, i.e. $\tau(\lambda) = 1/r_c(\lambda)$, which is compatible with what is expected in the non-corotating regions, close or beyond the light cylinder.
In this way, trajectories are twisted as well as curved, and thus require all three spatial dimensions to describe them.

\begin{table*}
    \centering
    \caption{Values or ranges of the relevant parameters entering in our calculation, together with a brief description of their function in the model.  }
    \begin{tabular}{ccc}\hline\hline
        Parameter & Function & Value/range selected \\\hline
        \textbf{Pulsar} \\
        $P$ & Period & Depending on the pulsar \\
        $\Dot{P}$ & Period derivative & Depending on the pulsar \\
        \hline
        \textbf{Spectral parameters} \\
        $E_{||}$ & Parallel electric field &  Best-fit value of an spectral fitting of the pulsar's high-energy SED \\
        $b$ & Magnetic gradient &  Best-fit value of an spectral fitting of the pulsar's high-energy SED \\
        $x_0$ & Lengthscale  & Best-fit value of spectral fitting of the pulsar's high-energy SED \\
        \hline
        \textbf{Map} \\
        $\Psi_{\Omega}$ & Inclination angle & [$9^{\circ}$, $18^{\circ}$, $27^{\circ}$, $36^{\circ}$, $45^{\circ}$, $54^{\circ}$, $63^{\circ}$, $72^{\circ}$, $81^{\circ}$, $90^{\circ}$] \\
        $\Delta \Psi_{\mu}$ & Meridional extent & [$5^{\circ}$, $10^{\circ}$, $15^{\circ}$] \\
        $\Delta R$ & Injection range  (in $R_{lc}$) & 0.5 \\
        \hline
    \end{tabular}
    \label{tab:fixed_parameters_model}
\end{table*}

\subsection{Emission directions of the traveling particles}

Due to the high Lorentz factor $\Gamma \gtrsim 10^6$ reached by the particles, the emission beam ($\propto \Gamma^{-1}$) is very narrow, so in our numerical scheme we consider the radiation emitted exactly in the direction of motion. The latter is determined as follows. At each $\lambda$, we consider a local particle coordinate system.
In this coordinate system, the instantaneous emission direction of a particle is $(\theta_e(\lambda), \phi_e(\lambda)) = (\alpha(\lambda), \chi)$, where $\chi$ is the gyration angle, $\chi \in [0, 2\pi]$.

We write the particle emission directions at each position in a pulsar coordinate system, in which the azimuthal coordinate $\phi_\Omega$ tracks the rotational phase of the star, and the meridional coordinate $\theta_{obs}$ corresponds to a given observer, i.e. is the so-called viewing angle. 
The transformation of the emission directions among the different coordinate systems, $(\theta_e(\lambda), \phi_e(\lambda))\rightarrow (\theta_{obs}, \phi_\Omega)$, is  calculated by a series of rotations considering the inclination angle (between the magnetic and rotation axes) $\Psi_\Omega$ and the time delay, as described in detail in \S~2 of \cite{light_curves_2019}. 
For the sake of clarity, note that we don't evolve the electromagnetic fields as in force-free or PIC simulations. We simply evolve the kinetic quantities ($\Gamma,\alpha$), defining them and the $r_c$, $B$ values in the inertial laboratory frame. We map the emission directions through these rotations which only change the coordinates. In other words, we don't consider different relativistic frames, we only change the coordinates we use to map directions in the sky, following the same concept explained in subsection 2.4 of \cite{Petri20}.

We construct the emission map by collecting the particles emission in a unit sphere centered at the star, in the pulsar coordinate system, and considering time delay.
For each particle, we can then consider the corresponding distribution of emission directions at each $\lambda$, $dD(\lambda)/d\Omega_{\Omega}$. Such distributions are simply circles with radius $\alpha$, thus approaching a single point in the sky as the particle accelerates.

In this study, we consider skymaps with $51\times102$ bins in $(\theta_{obs},\phi_\Omega)$, which make synthetic light curves having 102 phase bins, similar to the best-observed ones. 
We have now to specify how the coordinate $\lambda$ translates into the three-dimensional space, for a given location and shape of the accelerating region.

\subsection{Location in the magnetosphere and shape of the emission region}
\label{location_shape_emission_region}

We have taken into account to guide our definition of emission region the results of the latest magnetospheric PIC simulations \citep{Philippov18, Kalapotharakos18, Brambilla18, Hakobyan23, Kalapotharakos23}, all based on a magnetic dipole at the pulsar surface (thus giving the topology a plane of symmetry).
Their plots of current/plasma density and divergence of the electric field show that the likely emitting regions are located just outside the light cylinder, extending from the so-called Y-point where lines reconnect to the equatorial current sheet which completely surrounds the pulsar in the azimuthal direction. 
The Y-point is located at a magnetic colatitude $\Psi_\mu = \pi/2$ (being $\Psi_{\mu}$ the meridional angle from the magnetic axis).
Considering the outward extent, the regions have a shape similar to that of a ``ballerina skirt" (as dubbed by 
\citealt{review_electrodynamics_pulsar_magnetospheres}), i.e. having a magnetic colatitude which oscillates with the azimuthal angle and increases with the radial distance.
Such oscillations are maximum for an orthogonal rotator ($\Psi_\Omega=\pi/2$) and vanish for an aligned rotator ($\Psi_\Omega=0$).
An exact prescription of the shape of the emission region is not available to our knowledge. Analytical solutions for the split monopole \citep{Bogovalov99} provide ideal solution for the current sheet locations, but they are valid only far away from the light cylinder \citep{Tchekhovskoy16}. The shape we propose here is instead close to the Y-point, and it actually resembles qualitatively the innermost part of the wavy behaviour of the current sheet location in the meridional plane shown by Fig. 4 of \cite{Bogovalov99}. 

In order to take into account such qualitative characteristics, we define the emission region with a central value of the magnetic colatitude $\Psi_{\mu}^c$ given by the following function:
\begin{equation}
    \Psi_{\mu}^c (\xi_{\mu}, R, \Psi_{\Omega}) = \frac{\pi}{2} + A(R, \Psi_{\Omega}) \sin{(\xi_{\mu} - \pi/2)}~,
    \label{eq:psimu_function}
\end{equation}
where $\xi_\mu$ is the longitude measured in the coordinate system aligned with the magnetic dipole which, like $\Psi_\mu$, enter in the series of rotations mentioned above and described in \cite{light_curves_2019}, and $A$ represents the amplitude of the oscillations, which is effectively modeled as linear with $\Psi_\Omega$ and quadratic with the distance:
\begin{equation}
A(R, \psi_{\Omega}) = K ~ \Psi_{\Omega} ~ (R/R_{lc} - R^0/R_{lc})^2 .
\end{equation}
Here $R^0$ is the innermost point of the trajectory, which indicates the distance between the neutron star’s surface and the beginning of the emission region and is fixed to  $R^0 = 1 R_{lc}$.
The particular form of this formula, as well as the chosen and fixed value of $K=5$, qualitative represents the shape of the likely emitting regions (outer current sheets), as seen in the plots from PIC simulations in the cited literature. 
The region is located around $\Psi^c_{\mu} (\xi_{\mu}, R, \Psi_{\Omega})$, and has a meridional extent of $\Delta\Psi_{\mu}$, reaching magnetic colatitude values between $\Psi^c_{\mu} + \Delta\Psi_{\mu}/2$ and $\Psi^c_{\mu} - \Delta\Psi_{\mu}/2$.

We numerically sample the directions $\Psi_\mu$ (over the meridional extent $\Delta \Psi_\mu$) and $\xi_\mu$ (over $2\pi$) with a step of $\delta=0.88^{\circ}$, meaning that we consider a set of $(2\pi~\Delta\Psi_\mu)/\delta^2$ discrete particle trajectories.
With this resolution we are able to have well resolved emission maps, without relevant numerical features or noise in the light curves. The numerical convergence tests we have performed showed that the results remain basically unchanged when choosing a better resolution.

\begin{figure}
\centering
\includegraphics[width=0.50\textwidth]{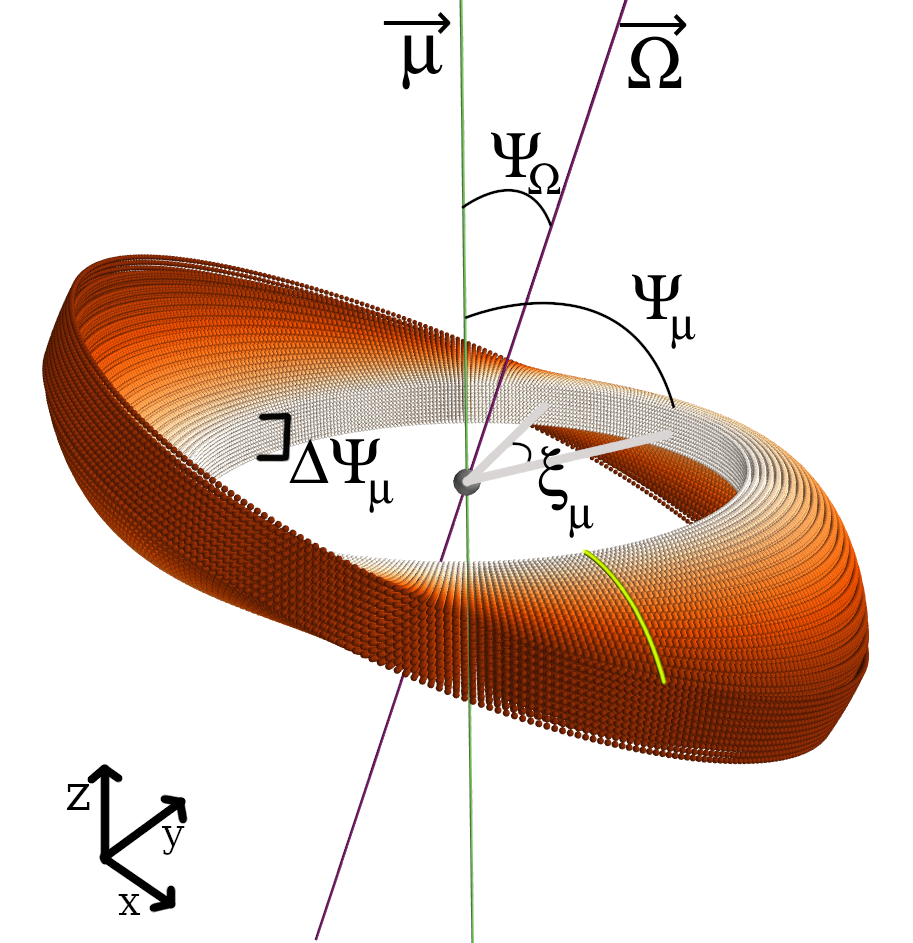}
    \caption{Geometrical representation of the magnetospheric emission region assumed in this work, for a pulsar for a geometry $\Psi_{\Omega} = 18^{\circ}$, $\Delta\Psi_{\mu} = 10^{\circ}$ and $\Delta R = 0.5$. This region is drawn here as the set of the gyration-average trajectories of the emitting particles attached to different rotating magnetic field lines. The yellow line is one of such trajectories. The gyration radius would be invisible in this scale. The color scale indicates the position $\lambda$ on the trajectory from the beginning of the region. The purple line is the rotational axis $\Vec{\Omega}$ and the green line is the magnetic axis $\Vec{\mu}$. The central gray sphere represents the position of the pulsar (its dimension is roughly at scale only if the period is a few milliseconds). The light gray lines (which represent a distance from the star of $R^0 = 1.0 R_{lc}$) that connect the star and the base of the region are contained in the plane perpendicular to the magnetic axis.
    }
    \label{fig:plot_region_with_angles}
\end{figure}

We show the resulting shape of the emitting region in Fig.~\ref{fig:plot_region_with_angles}, for an arbitrary geometry.
The parameters $\Psi_{\Omega}$ and $\Delta\Psi_{\mu}$ are free geometrical parameters of our model. 
We remark the difference between these two parameters: while the former is the widely used inclination angle, the latter is an effective parameter of our model to quantify the meridional extent of the region.

Another physical parameter is required to fully describe the location of the region: the distance of the injection points, where particles start traveling in the region. We parametrize this distance with the variable $x_{in}$, see e.g., \cite{diego_solo,systematic_2019}.
In \cite{sc_emitting_regions_2022} we showed that the spectra of high-energy pulsars remained basically unchanged when considering several $x_{in}$'s, instead of a single one. However, the extension of the region affects the skymap, as we will see later on.
Here, accordingly, we have considered several injection points $x_{in}$ (in practice, 21, see Appendix \ref{app:numerical_convergence_tests}).
These are equally distributed from $1~ R_{lc}$ 
to $1.5~ R_{lc}$, i.e. an injection range $\Delta R=0.5 ~R_{lc}$, in which accelerating particles are injected and begin their trajectories. 
We thus have an extended injection of particles all throughout this region.
Notice that in this way, the region starts at a distance from the star of $R = 1.0 R_{lc}$, i.e. at the light cylinder.
We remark that this is our default configuration, but our model is flexible 
in this regard and we can consider different radial extents of the region and different number of injection points.

Once we have the three-dimensional description, we can associate each injection place and following values of $\lambda$ to a specific point in space, so that for each point of a trajectory starting at a given $x_{in}$, we can calculate the corresponding emission directions, via the distribution $dD(\lambda)/d\Omega_{\Omega}$, following the procedure described in \cite{light_curves_2019}.

\subsection{Radiation emitted from the emission region}
\label{sc_radiation}

So far, we presented only the geometrical part of the model, with no reference to the emission of the particles that move along the emission region.
We use our radiative model based on the trajectory explained above and the expected synchro-curvature emission of the particles (see \citealt{cheng_zhang, compact_formulae}) at each location. The model computes the synchro-curvature radiation emitted by the particles at each position of their trajectory, $dP_{sc}(\lambda)/dE$.
Our radiative model, based on a critical review of the existing gap models \citep{outer_gap_model_paper_1,outer_gap_model_paper_2}, has successfully fit the phase-averaged spectra of the entire $\gamma$-ray pulsar population \citep{Vigan_2015b}.
Our model was later extended to deal with the X-ray range too \citep{diego_solo} 
and improved the description of the injection region and used as a tool for possible period inferences
\citep{systematic_2019,sc_emitting_regions_2022,period_inference}.
For the spectral description here we take the latest incarnation of the model \citep{sc_emitting_regions_2022} with improvements regarding the injection and particle density description.

In brief, the spectral modeling includes the trajectory calculation mentioned above and has only three free parameters, $E_{\parallel}$, $b$ and $x_0$, which completely determine the spectral shape for a given pulsar (i.e., a given $P$ and $\dot{P}$), plus a normalization $N_0$. The latter is used only to fit the total observed flux (assuming a distance), and doesn't affect the spectral shape or the skymap, therefore it has no relevance in the present study.

The parallel electric field $E_{\parallel}$ and the magnetic gradient $b$ alone define the trajectories that the particles travel in the emitting regions, as mentioned above.
The parameter $x_0$ encompasses all physical effects not considered in our effective approach, such as a non-constant parallel electric field or the presence of particles traveling backwards in the region. 
It is included in a relative weight given to the particle distribution ($dN/d\lambda = N_0 e^{-(\lambda - \lambda_{in})/x_0}/[x_0(1 - e^{-(\lambda_{out} - \lambda_{in})/x_0})] $), so that the initial parts of the trajectories (having large pitch angles and X-ray dominating emission) have more relative weight.
Here $\lambda_{in}$ and $\lambda_{out}$ are the initial and last positions of the trajectory.

To keep our effective approach simple for the time being, we consider that the radiation produced by the particles at each rotating magnetic field line and the effective particle distribution in it is the same in all the lines of the region.
Therefore, convolving the single-particle radiated power $dP_{sc}(\lambda)/dE$ with the mentioned weights, we obtain the radiation emitted 
\begin{equation}
    \frac{dP_{tot}}{dE} = \int_{\lambda_{in}}^{\lambda_{out}} \left\langle\frac{dP_{sc}(\lambda)}{dE}\right\rangle\frac{dN(\lambda)}{d\lambda} d\lambda~.
    \label{eq:convolved_power_spectra}
\end{equation}
In the latter integral, as commented above, we keep into account the recently introduced treatment of emission coming from an extended emission, as explained in Sec. 2.5 of \cite{sc_emitting_regions_2022}.
This theoretical spectral energy distribution (SED) represents the total emission of all particles in a rotating field line per unit energy, and can be fitted to an observational spectra, as we did in our previous works.

\subsection{Emission skymap}
\label{emission_map}

The skymap for a given set of spectral parameters is obtained by considering all the lines, the injection points and integrating over the solid angle the distribution of directions. The final expression for the energy-dependent emission skymap is then:
\begin{eqnarray}
    M_E(\theta_{obs}, \phi_{\Omega},E) &=& \int_{R^0}^{R^0+\Delta R}\int^{\lambda_{out}}_{\lambda_{in}} \int_{0}^{2\pi}\int_{\Delta\Psi_{\mu}}  \nonumber \\
        &&
    \hspace{2.1cm}
    \frac{dD(\lambda)}{d\Omega_{\Omega}} \left\langle\frac{dP_{sc}(\lambda)}{dE}\right\rangle \frac{dN(\lambda)}
    {d\lambda} 
    \nonumber \\
    &&
    \hspace{2.15cm}
    d\Psi_{\mu} \, d\xi_{\mu} \, d\lambda \, dx_{in}~,
    \label{eq:emission_map}
\end{eqnarray}
which represents the photon flux per unit energy, per unit solid angle, emitted by particles injected all along the region. 
We shall consider that the final map is normalized to the total emission, i.e. to the summed emission of the whole map, therefore being insensitive on the normalization $N_0$ used in the spectral fit.

Such a skymap represents the projection over a sphere at infinity of the radiation emitted by the whole population of accelerated particles. For a given observer with viewing angle $\theta_{obs}$, the light curve (intensity as a function of the phase) is  given by an azimuthal cut on the skymap, integrating $M_E$ over a given energy range. In this study we focus on the Fermi-LAT $\gamma$-ray band, 100 MeV -- 300 GeV.
Table \ref{tab:fixed_parameters_model} shows the values for the relevant free parameters of the spectral+geometrical model.

\section{Synthetic skymaps and light curves of selected pulsars}
\label{SMs_LCs}

\begin{table*}
    \centering
    \caption{Timing and best-fitting spectral parameters of the ten selected pulsars, The spectral model here considers 21 injection points evenly distributed from 1.0 to 1.5 $R_{lc}$.
    Last two columns show the number of peaks of the observational light curve released in the 2PC and 3PC for each pulsar.}
    \begin{tabular}{crcccccccc}
        \hline\hline
        Pulsar & P [ms] & $\Dot{P}$ [s/s] & $\log{E_{\parallel}}$ [V/m] & $\log{(x_0/R_{lc})}$ & $b$ & $\log{N_0}$ & $B_{lc}^{b=3} (G)$ & Peaks 2PC & Peaks 3PC \\\hline
        J0007+7303 & $315.9$ & $3.57 \times 10^{-13}$ & $8.29^{+0.01}_{-0.01}$ & $-2.67^{+0.01}_{-0.01}$ & $2.49^{+0.02}_{-0.02}$ & $33.95$  & $6.28 \times 10^3$ & 2 & 3 \\ 
        J0205+6449 & $65.7$ & $1.92 \times 10^{-13}$ & $8.98^{+0.01}_{-0.01}$ & $-3.08^{+0.01}_{-0.01}$ & $2.62^{+0.02}_{-0.02}$ & $34.02$ & $2.32 \times 10^5$ & 2 & 2 \\
        J0218+4232 & $2.3$ & $7.74 \times 10^{-20}$ & $9.27^{+0.02}_{-0.01}$ & $-2.43^{+0.01}_{-0.01}$ & $2.68^{+0.05}_{-0.05}$ & $33.90$ & $6.32 \times 10^5$ & 2 & 1 \\
        J0633+1746 & $237.1$ & $1.10 \times 10^{-14}$ & $7.66^{+0.01}_{-0.01}$ & $-1.92^{+0.01}_{-0.01}$ & $2.25^{+0.02}_{-0.01}$ & $32.35$ & $2.26 \times 10^3$ & 2 & 3 \\
        J0835\textminus4510 & $89.4$ & $1.25 \times 10^{-13}$ & $8.24^{+0.01}_{-0.01}$ & $-2.38^{+0.01}_{-0.01}$ & $2.87^{+0.02}_{-0.02}$ & $33.77$ & $8.73 \times 10^4$ & 3 & 4 \\
        J1513\textminus5908 & $151.6$ & $1.53 \times 10^{-12}$  & $7.31^{+0.01}_{-0.01}$ &  $-2.42^{+0.01}_{-0.01}$  & $3.23^{+0.01}_{-0.01}$ &  $37.51$  & $8.19 \times 10^4$ & 1 & 1 \\
        J1809\textminus2332 & $146.8$ & $3.44 \times 10^{-14}$ & $8.25^{+0.01}_{-0.01}$ &  $-2.56^{+0.01}_{-0.01}$  & $2.58^{+0.03}_{-0.02}$  & $33.77$ & $1.32 \times 10^4$ & 2 & 3 \\
		J2021+3651 & $103.7$ & $9.56 \times 10^{-14}$ & $8.51^{+0.01}_{-0.01}$  & $-2.78^{+0.01}_{-0.01}$  &  $2.95^{+0.01}_{-0.01}$ &  $34.82$ & $5.26 \times 10^4$ & 2 & 3 \\
		 J2021+4026 & $265.3$ & $5.42 \times 10^{-14}$ & $7.97^{+0.01}_{-0.01}$ &  $-2.51^{+0.01}_{-0.01}$    &$2.78^{+0.05}_{-0.04}$  & $35.03$ & $3.78 \times 10^3$ & 2 & 3 \\
		 J2229+6114 & $51.7$ & $7.79 \times 10^{-14}$ & $8.56^{+0.01}_{-0.01}$ &  $-2.55^{+0.01}_{-0.01}$   & $2.70^{+0.01}_{-0.01}$ &  $34.28$ & $2.72 \times 10^5$ & 2 & 2 \\
        \hline
    \end{tabular}
    \label{tab:parameters_pulsars}
\end{table*}

\subsection{Set of geometries and sample of pulsars}

Consider now a given pulsar, with certain timing properties $P$, $\Dot{P}$, and a given set of the three parameters ($E_{\parallel}$, $x_0$, $b$) that best fits the X and $\gamma$-ray observed spectral energy distributions, following our previous works \citep{sc_emitting_regions_2022}.
The spectral model alone, however, doesn't take into account the spatial distribution of the emitted radiation: it gives just the total radiation emitted, integrated over the entire skymap.
We now focus on studying the variety of observed light curves, for different geometrical parameters. 
The inclination angle $\Psi_{\Omega}$ can vary from case to case, since it is still not clear how it evolves \citep{Philippov14_time_evolution_inclination_angle}. 
We will also consider different values of our free parameters $\Delta R$ and $\Delta\Psi_{\mu}$.
For each combination of realistic values of $\Psi_{\Omega}$ and $\Delta\Psi_{\mu}$ we build a skymap, from which different light curves are obtained when varying $\theta_{obs}$.
Realistic here is used to imply that the region generated will have a shape in agreement with those obtained with PIC simulations and at the same time it is not presenting any physical inconsistency, such as very tiny or very large regions or overlapping layers.
Note that the rotational phases $0^{\circ}$ and $180^{\circ}$ of the skymaps correspond to the magnetic poles, i.e. to the plane containing the rotational and magnetic axes of the pulsar.

To span the possible outcome of our model we consider 10 values of the inclination angle, $\Psi_{\Omega} \simeq [9^{\circ}, 18^{\circ}, 27^{\circ}, 36^{\circ}, 45^{\circ}, 54^{\circ}, 63^{\circ}, 72^{\circ}, 81^{\circ}, 90^{\circ}]$; and 3 for the meridional extent of the base of the emission region, $\Delta\Psi_{\mu} \simeq [5,^{\circ}, 10^{\circ}, 15^{\circ}]$. 
With our particular choice of the grid of the parameters, we have a total of 30 different geometries.
For each corresponding skymap we consider 51 different types of observers, covering from the North to the South poles.
Due to discretization, such observers are indeed a bin in the skymap latitude, with a thickness of $\pi/50 \simeq 3.5^{\circ}$ centered on a value of $\theta_{obs}$. 
Thus, each light curve represents the radiation collected inside that bin. 
The values of $\theta_{obs}$ we will give later correspond to the central value of each bin.

In this way, we end up with a set of 1530 different light curves for a given pulsar
[10 (inclination) $\times$ 3 (meridional extent of emission region) $\times$ 51 (observers)].
Notice that this number is given solely by our choice of the number of geometries and observers, which is a compromise between computational cost and sufficient size of the synthetic sample. For a given pulsar, the computational time required is typically a few hours.

We aim at studying the variety of obtained light curves (in general and as a function of parameters) by looking at how many observers detect light and which features the curves have.
To this purpose, we apply a peak recognition algorithm to each of the light curves.
The details of the algorithm are shown in Appendix \ref{app:peak_recognition_algorithm}.
Using it, we can thus extract basic information from each light curve, such as the number of peaks, as well as the width, intensity and location (rotational phase) of each peak. 

Here we consider as a first example ten pulsars having good {\it Fermi}-LAT data both for their spectra and light curves. 
The ten pulsars are: J0007+7303 (a young pulsar), J0205+6449 (a bright young pulsar), J0218+4232 (a millisecond pulsar), J0633+1746 (Geminga), J0835-4510 (Vela, the brightest $\gamma$-ray source in the sky), J1513-5908 (one of the few pulsars detected at MeV energies), J1809-2332, J2021+3651, J2021+4026, J2229+6114 (four young pulsars as well). 
Table \ref{tab:parameters_pulsars} shows their timing and spectral parameters.
They are representative of the variety of the $\gamma$-ray pulsars population in terms of their timing properties, the number of peaks of their light curves in the Second and Third Fermi Catalogs of $\gamma$-ray Pulsars (\citealt{2fpc,3fpc}, hereafter 2PC and 3PC, respectively) and the values of their best-fit spectral parameters, for instance.

\subsection{Effects of the geometry on the skymaps and light curves}\label{effects_different_geometries_sms_lcs}

\begin{figure*}
    \centering
    \includegraphics[width=0.33\textwidth]{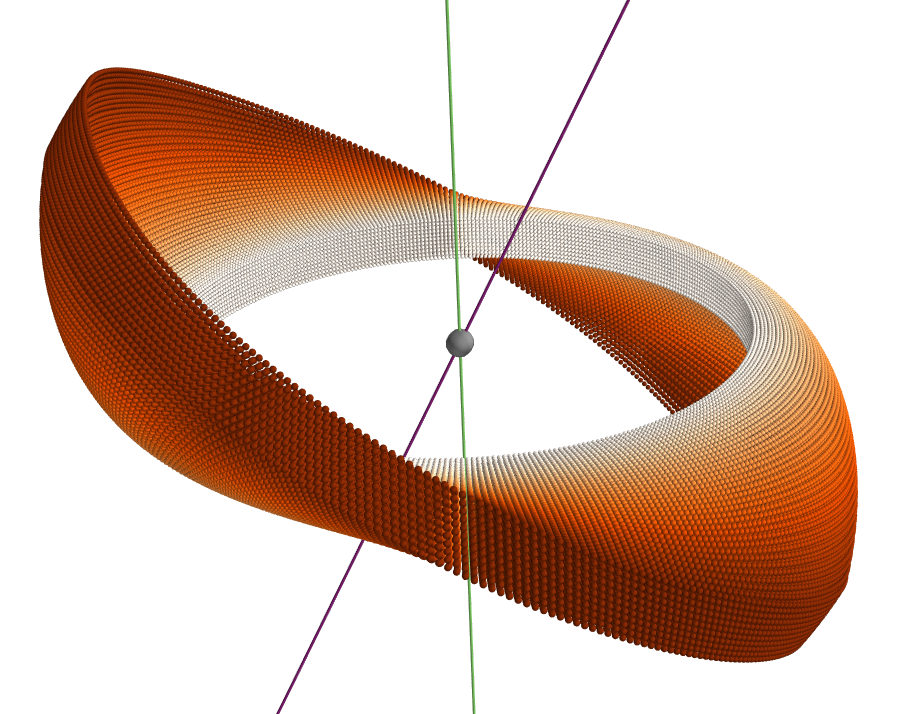}%
    \includegraphics[width=0.35\textwidth]{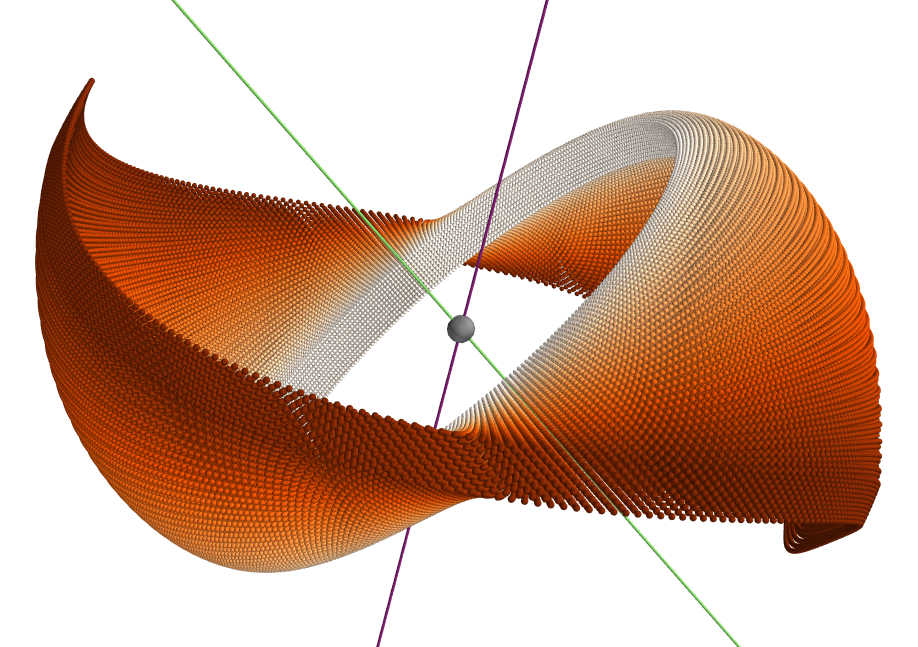}%
    \includegraphics[width=0.30\textwidth]{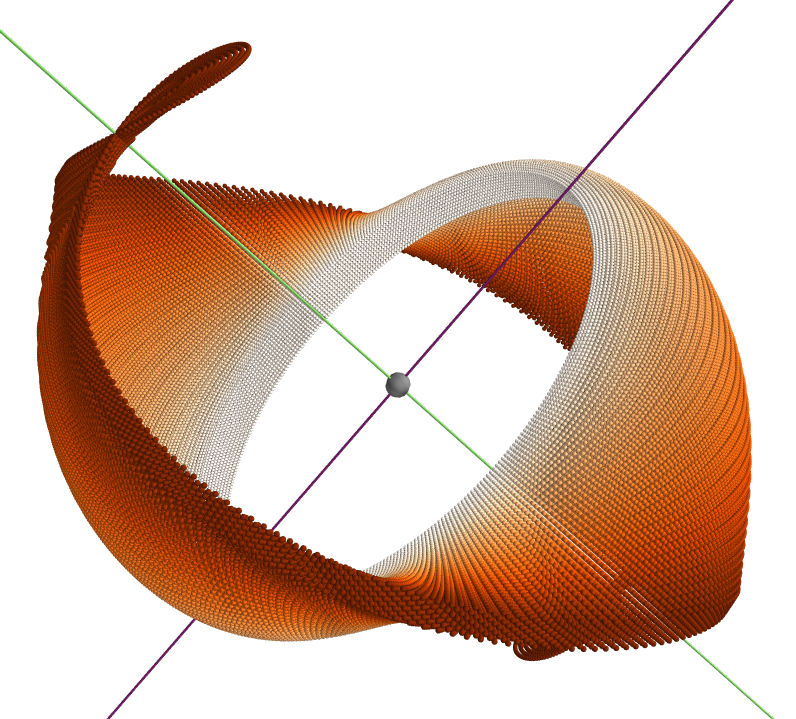}\\
    \includegraphics[width=0.33\textwidth]{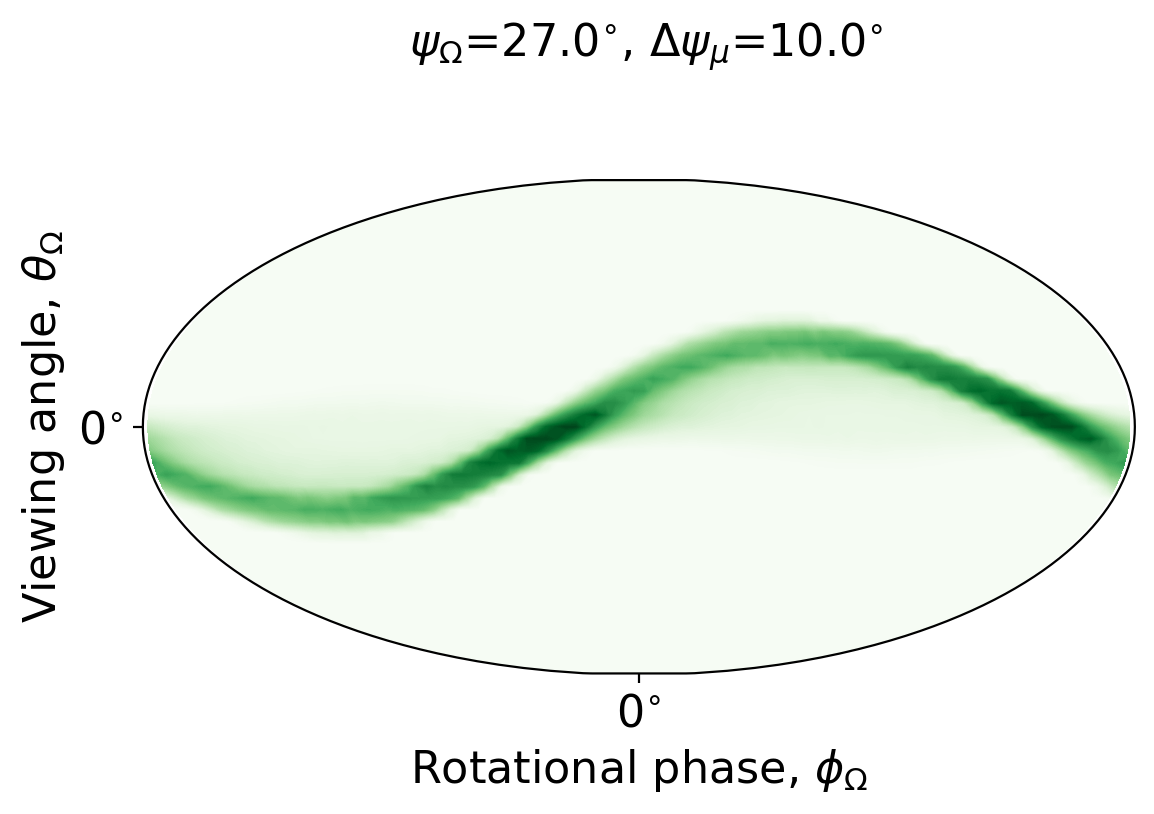}%
    \includegraphics[width=0.33\textwidth]{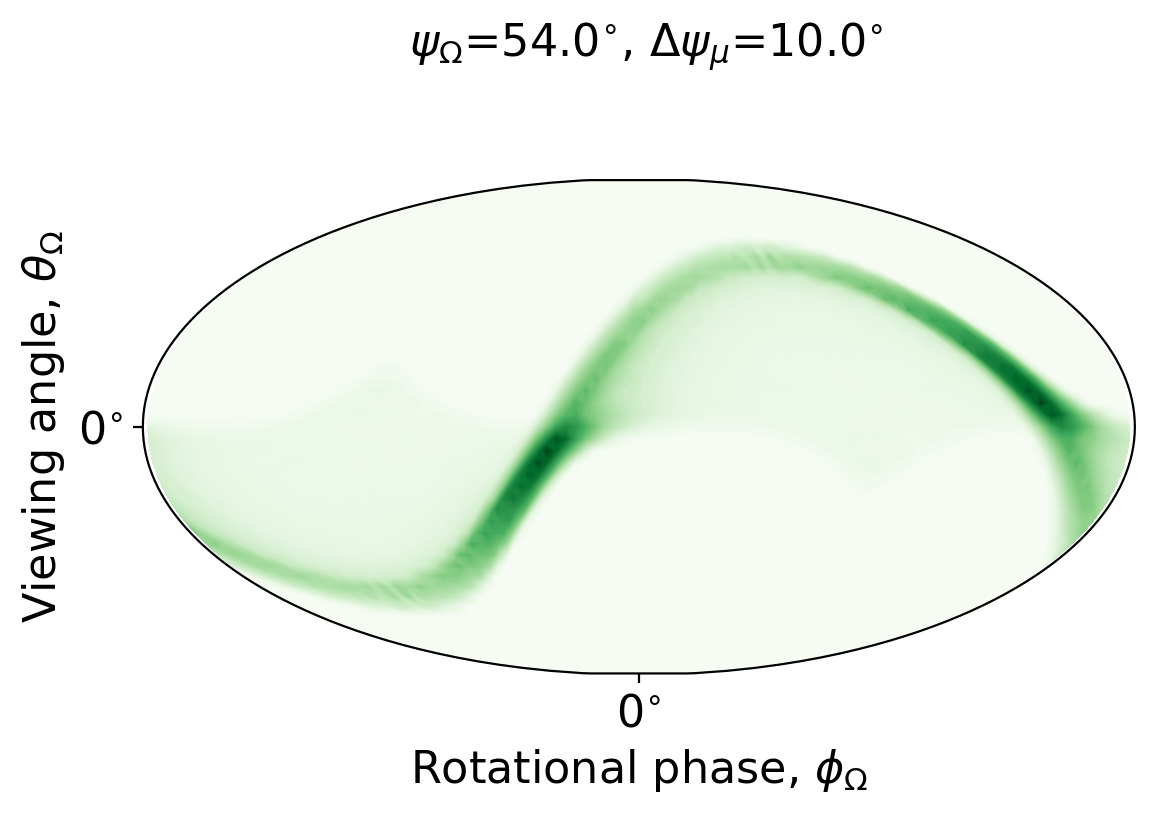}%
    \includegraphics[width=0.33\textwidth]{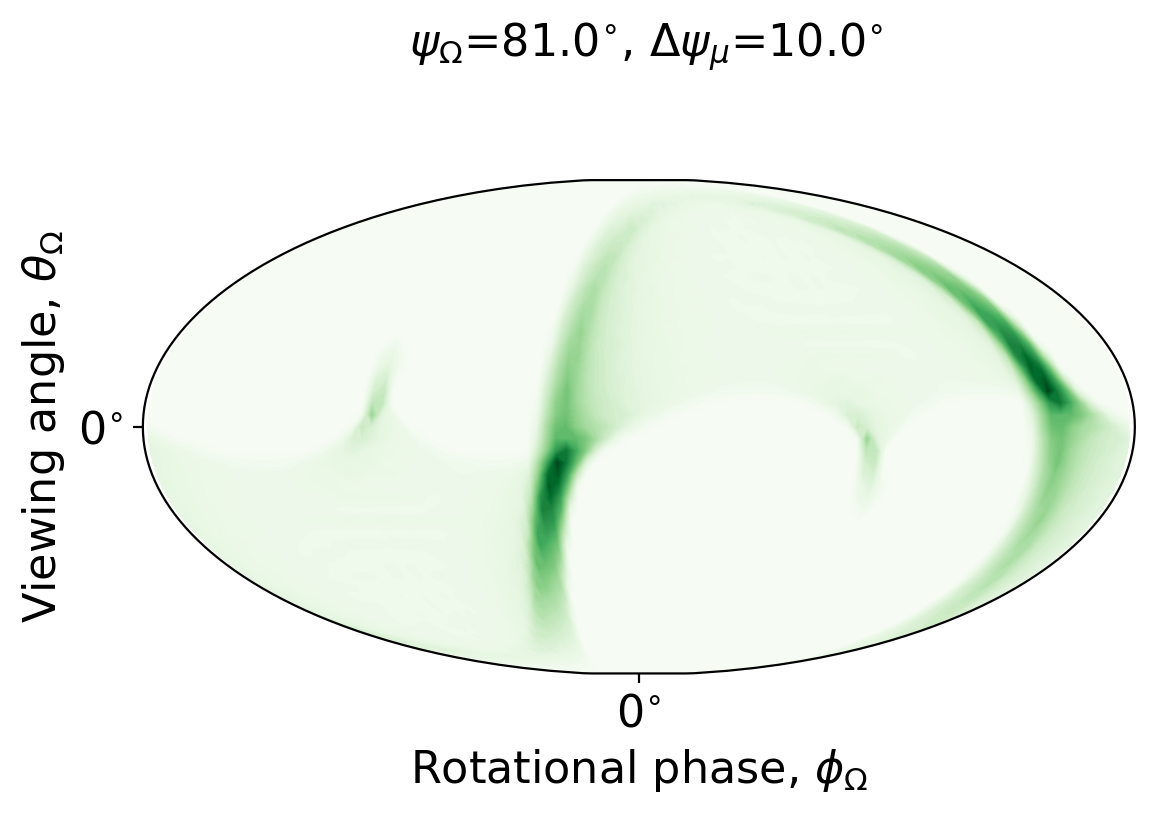}\\
    \includegraphics[width=0.33\textwidth]{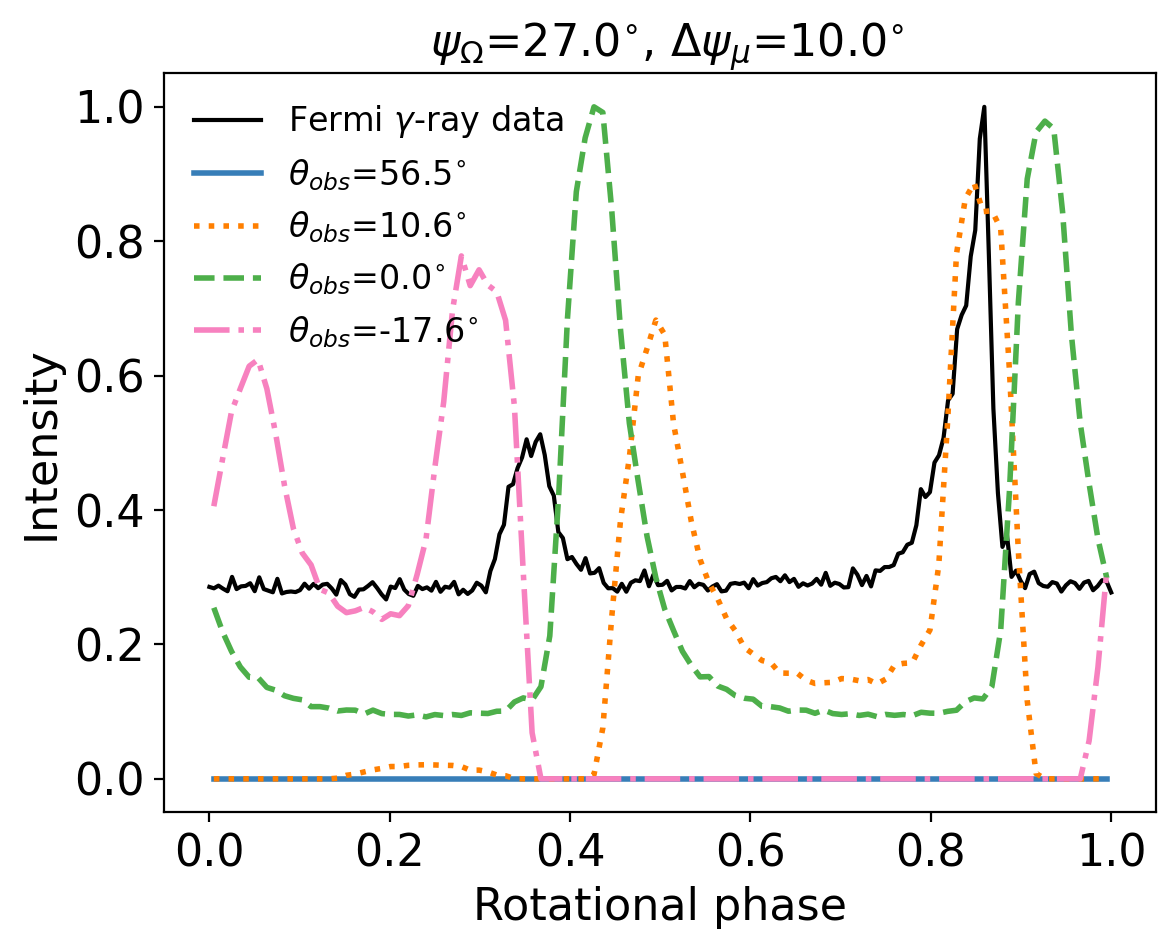}%
    \includegraphics[width=0.33\textwidth]{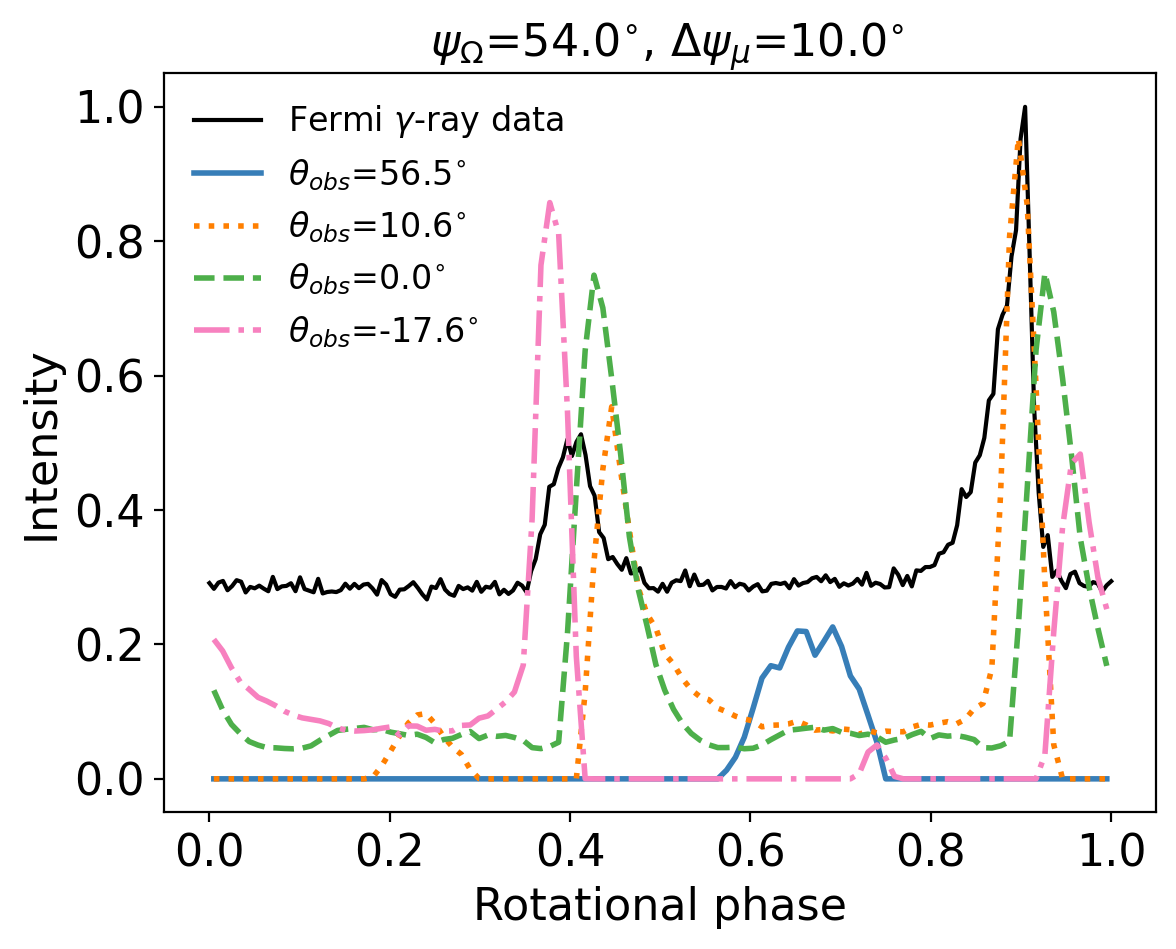}%
    \includegraphics[width=0.33\textwidth]{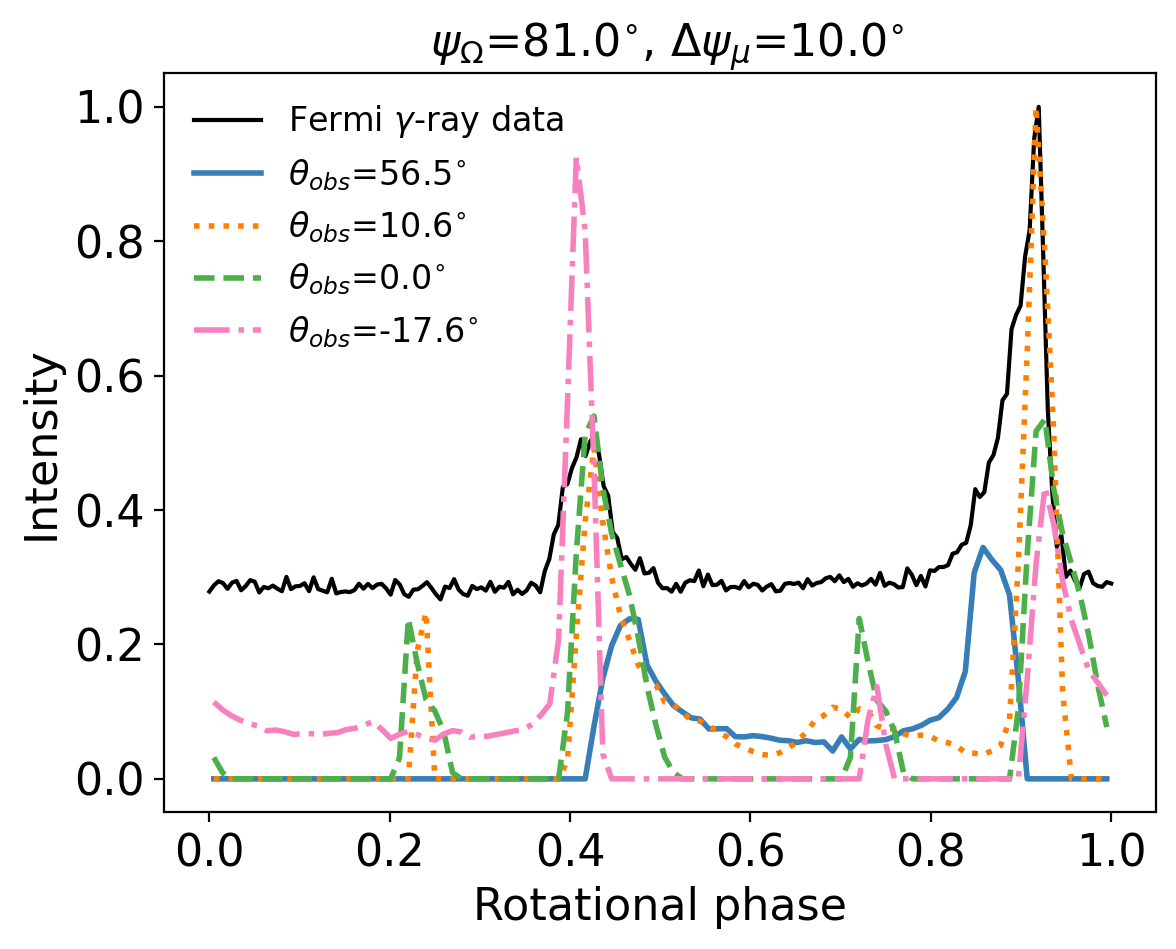}\\
    \includegraphics[width=0.33\textwidth]{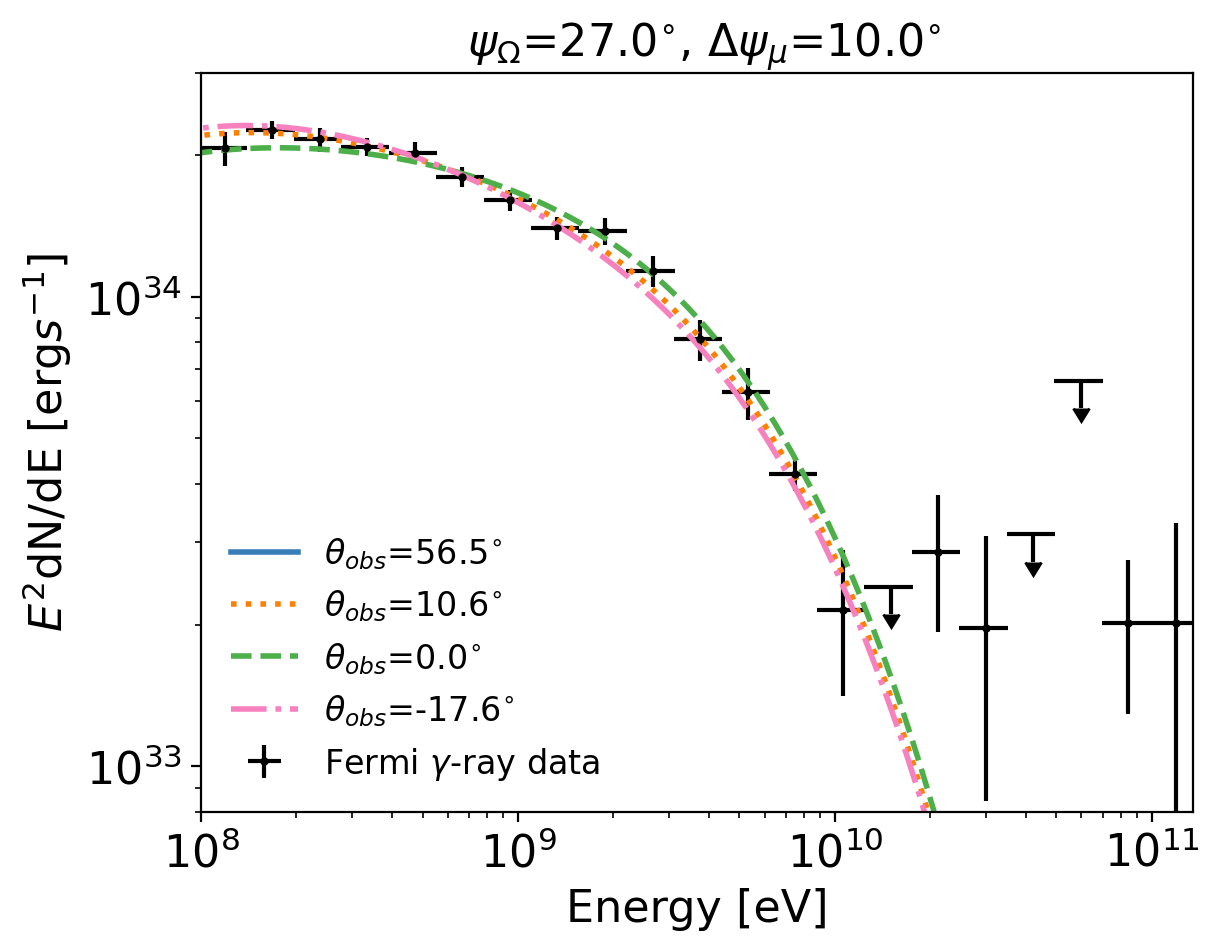}%
    \includegraphics[width=0.33\textwidth]{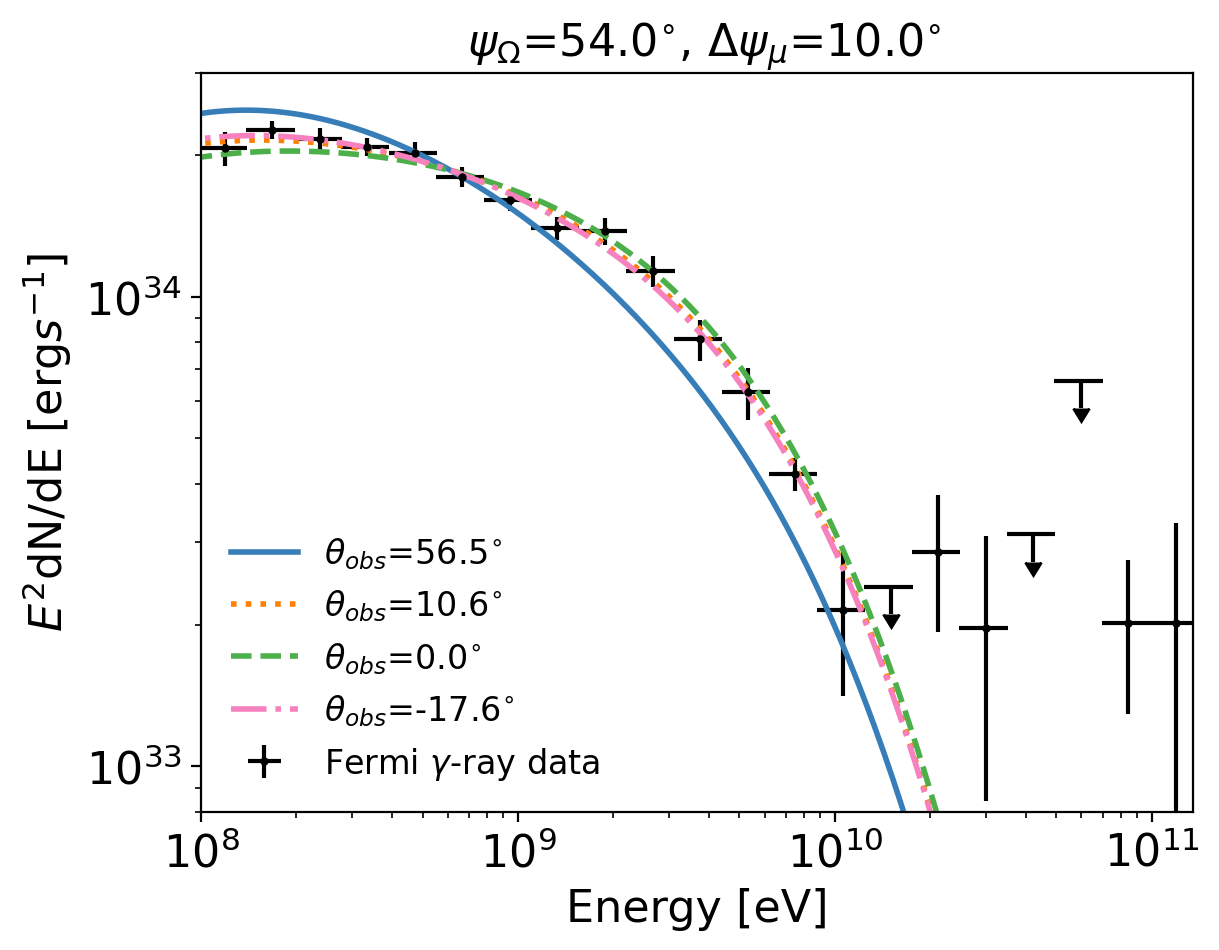}%
    \includegraphics[width=0.33\textwidth]{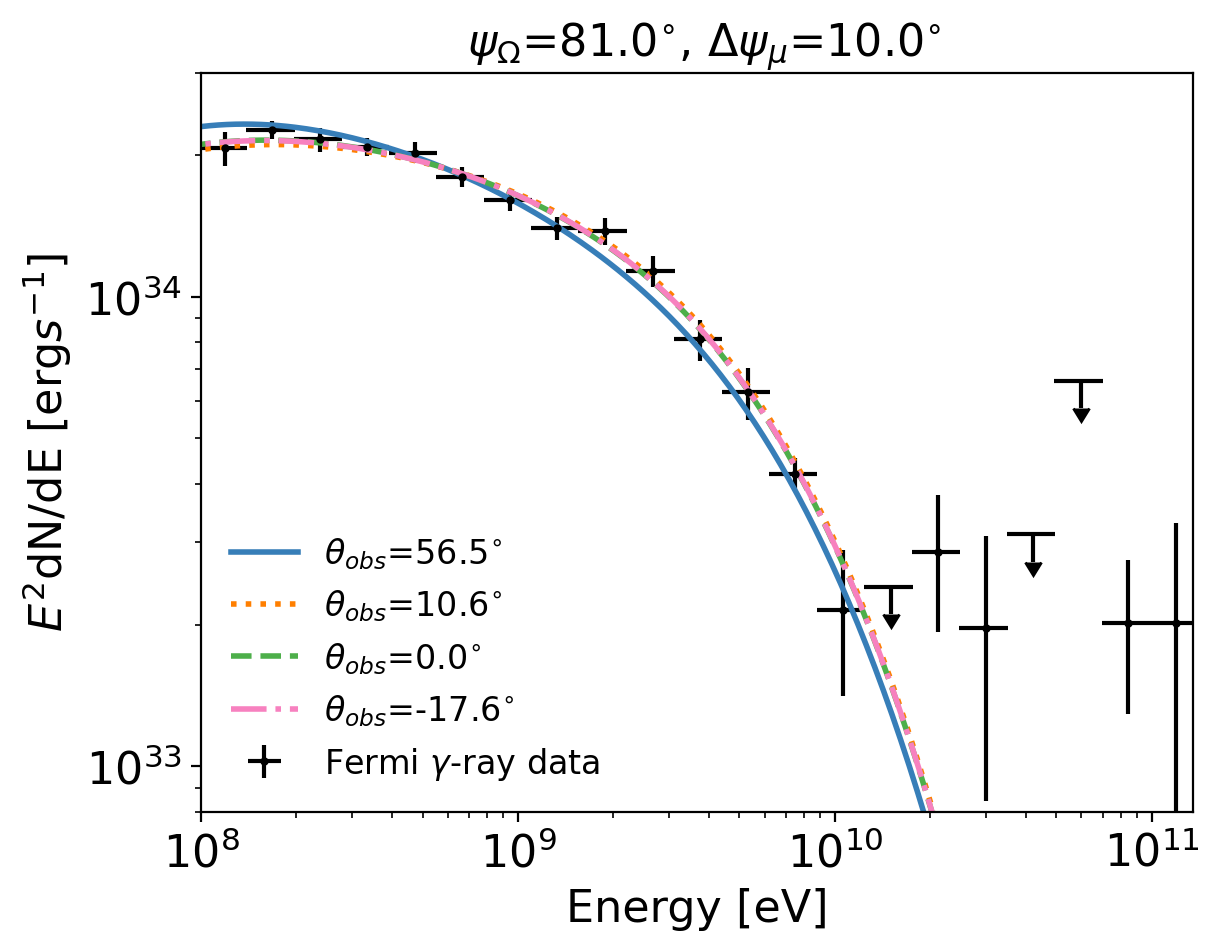}\\
    \caption{First row: representative examples of the geometry of the magnetospheric emission region of the pulsar J0205+6449 for different values of the inclination angle $\Psi_{\Omega}=\{27^\circ,54^\circ,81^\circ\}$ (increasing from left to right). The meridional extent of the region is $\Delta\Psi_{\mu}=10^\circ$ and the injection range $\Delta R=0.5~R_{lc}$. Second row: corresponding synthetic emission maps. These skymaps (as well as their corresponding light curves) are obtained by integrating $M_E$ in energies, in the $\gamma$-ray band, 100 MeV -- 300 GeV. Third row: corresponding synthetic light curves (in colors) for a few, arbitrarily chosen, observers $\theta_{obs}$. Notice that light curves are not normalized at the maximum intensity, which is only done for peak classification, but at the maximum intensity of the emission map, just for visualization and comparison purposes. Black lines correspond to the observational light curve of J0205+6449 seen by the \emph{Fermi} telescope and released in the 3PC, normalized to its maximum intensity. For simplicity it is arbitrarily aligned with the light curve seen by the observer at $\theta_{obs}=10.6^{\circ}$ (we could also rotate the synthetic light curves in phase in order to compare them with the observational one). Recall that producing a light curve fitting is not our goal here, we show this just for qualitative comparison of the real case with results coming from an assumption of fixed values of the parameters.
    Fourth row: corresponding $\gamma$-ray spectra seen by the same observers, showing differences in shape, together with observational $\gamma$-ray data from \emph{Fermi}.
    }
\label{fig:skymaps_0205_varying_psiomega}
\end{figure*}

\begin{figure*}
    \centering
    \includegraphics[width=0.33\textwidth]{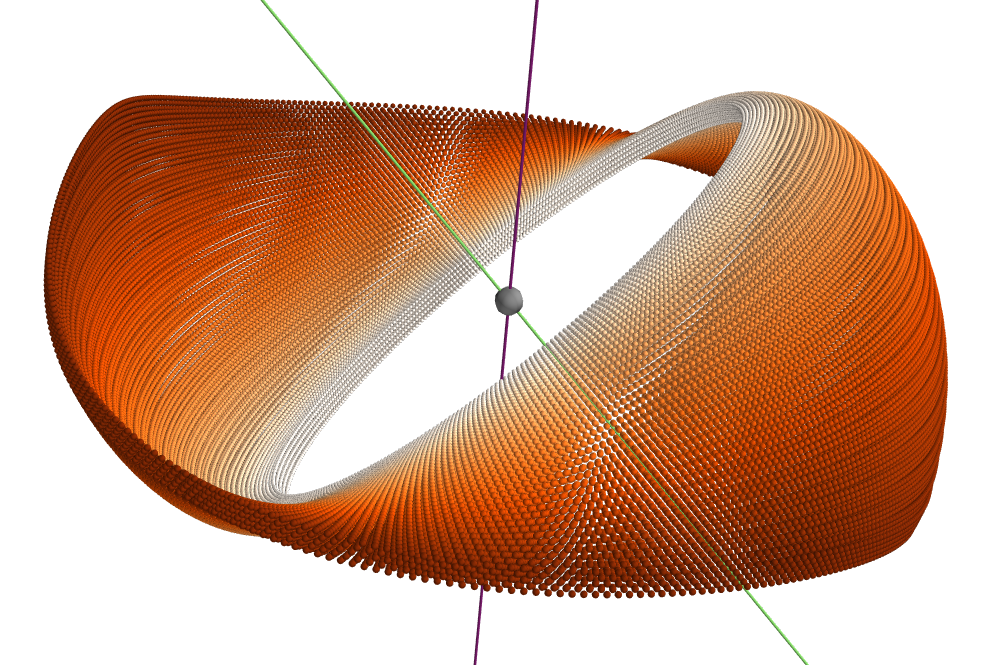}%
    \includegraphics[width=0.33\textwidth]{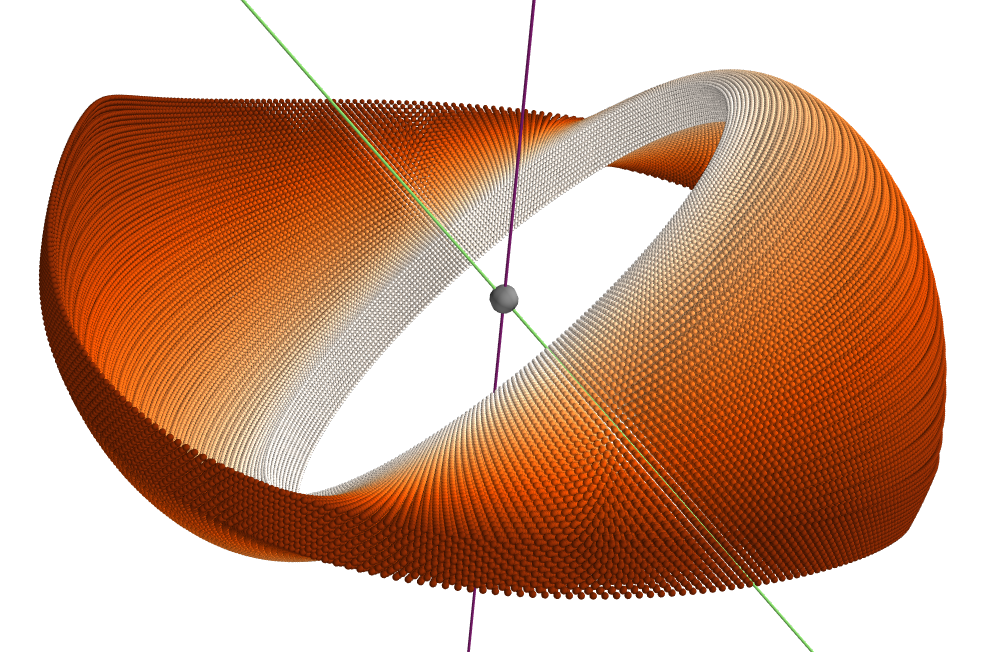}%
    \includegraphics[width=0.33\textwidth]{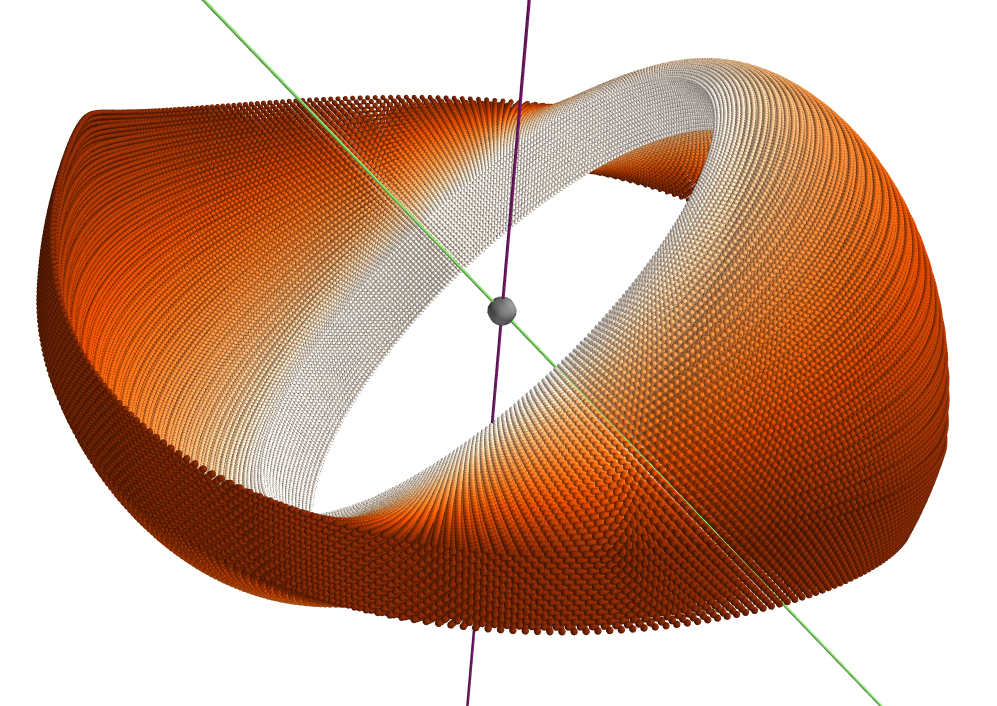}\\
    \includegraphics[width=0.33\textwidth]{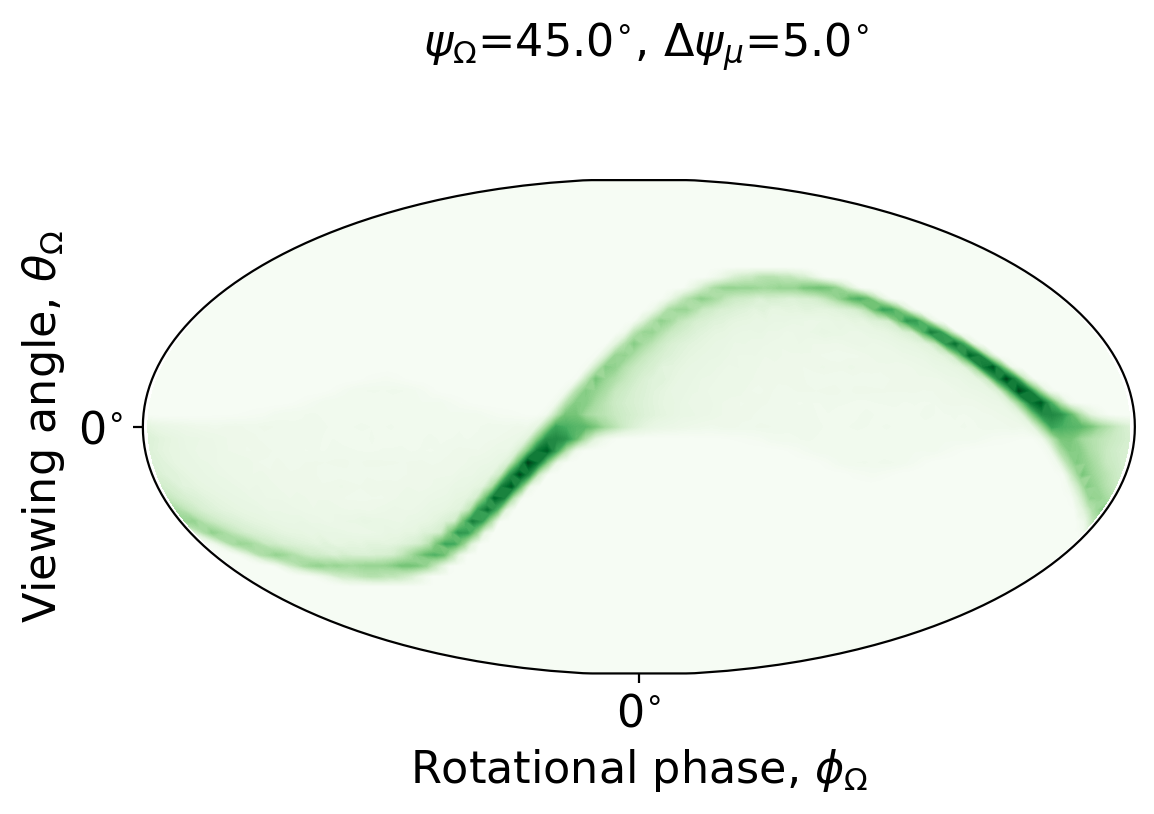}%
    \includegraphics[width=0.33\textwidth]{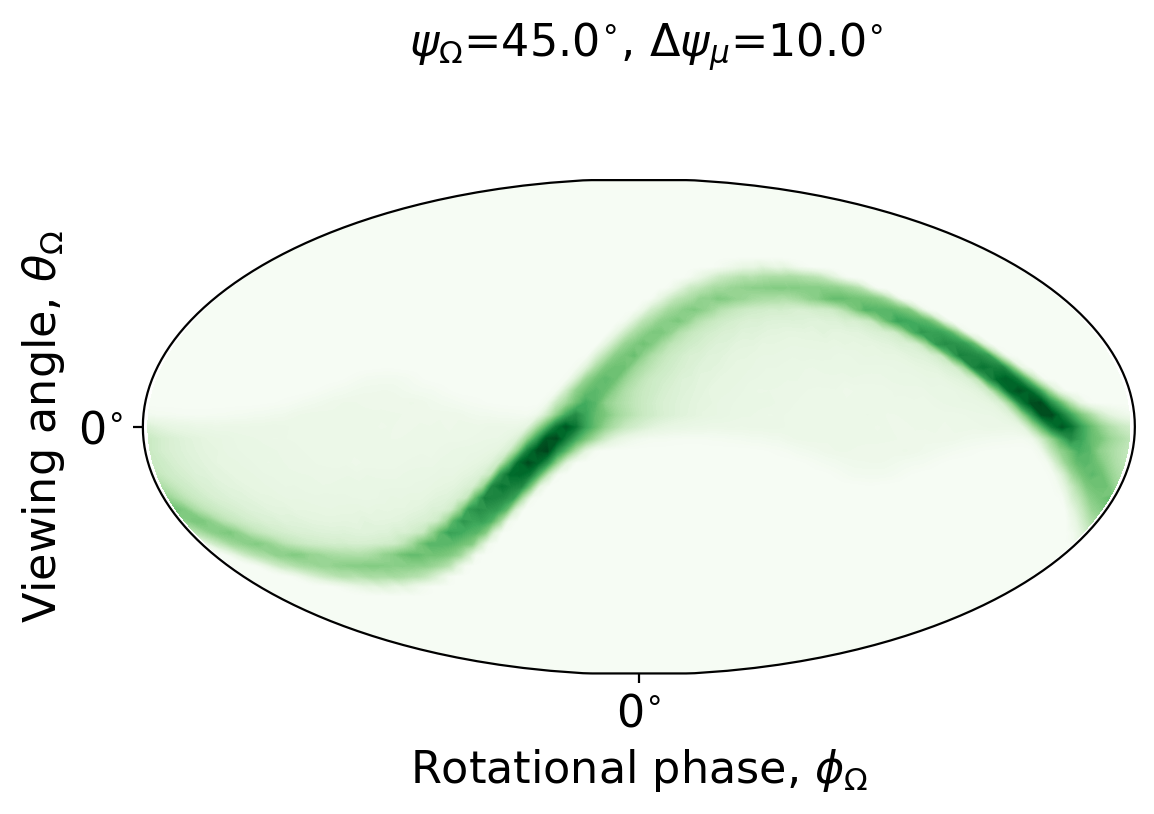}%
    \includegraphics[width=0.33\textwidth]{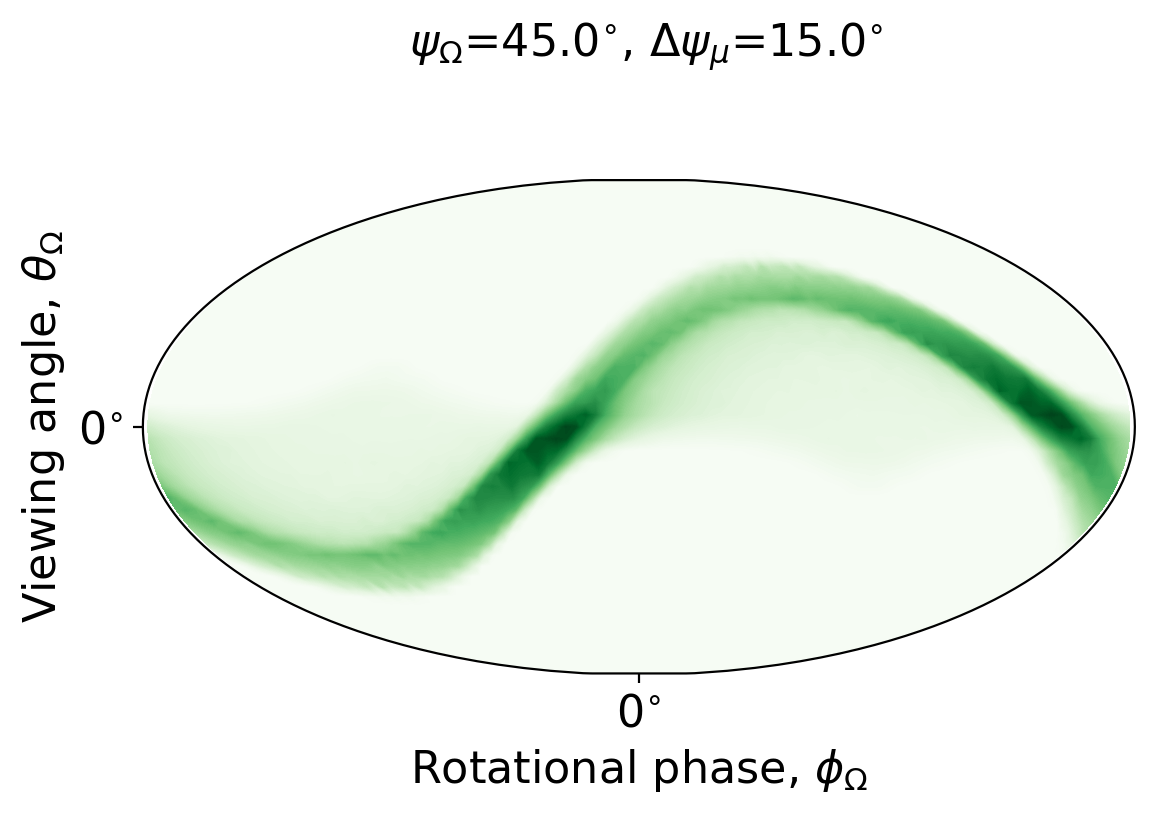}\\
    \includegraphics[width=0.33\textwidth]{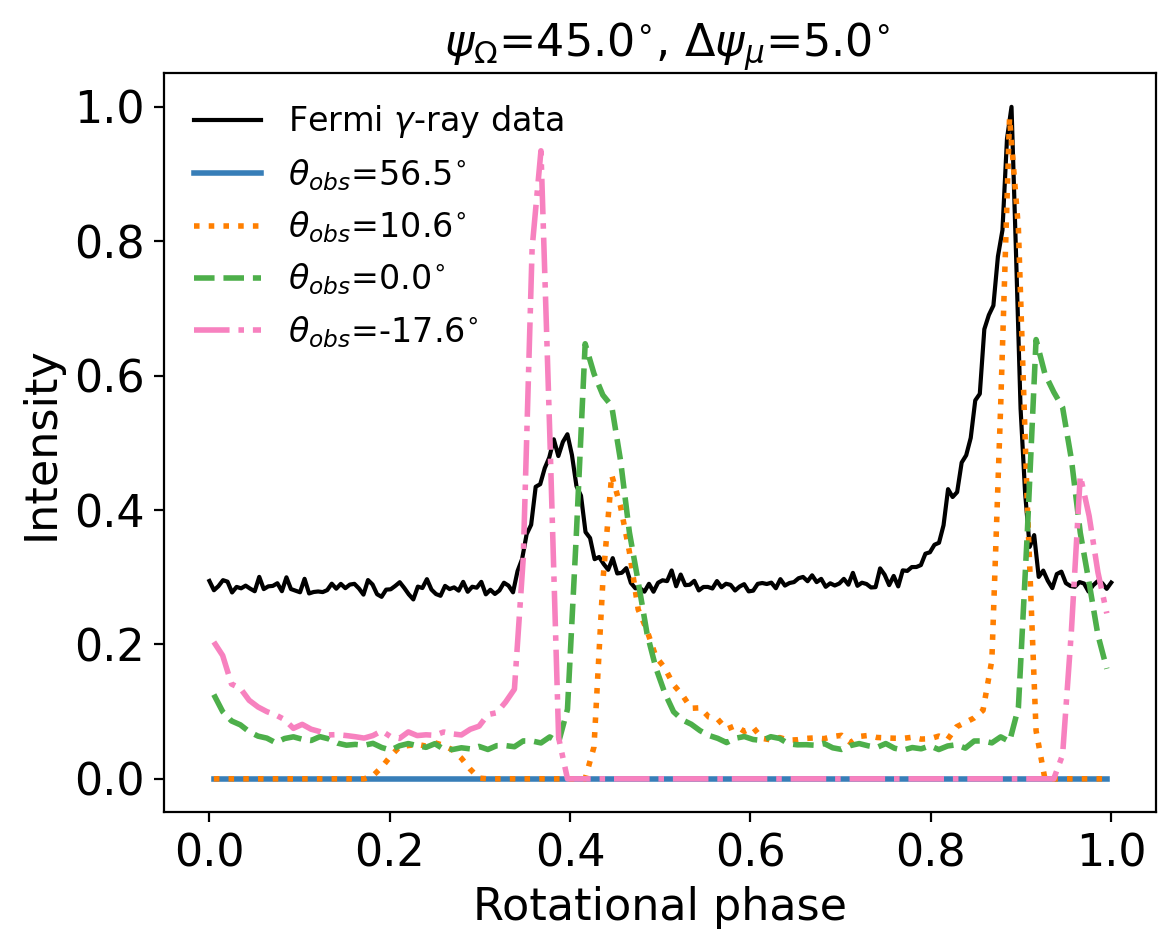}%
    \includegraphics[width=0.33\textwidth]{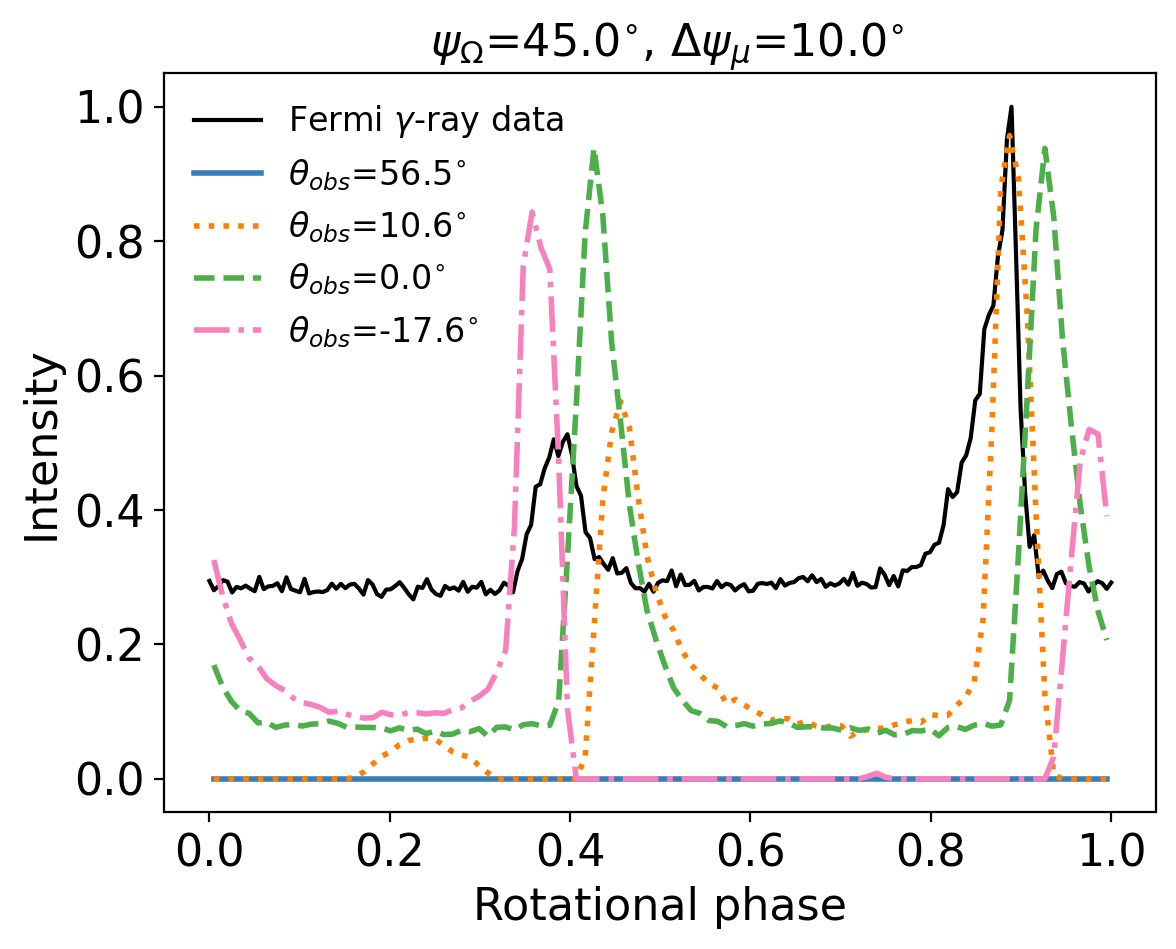}%
    \includegraphics[width=0.33\textwidth]{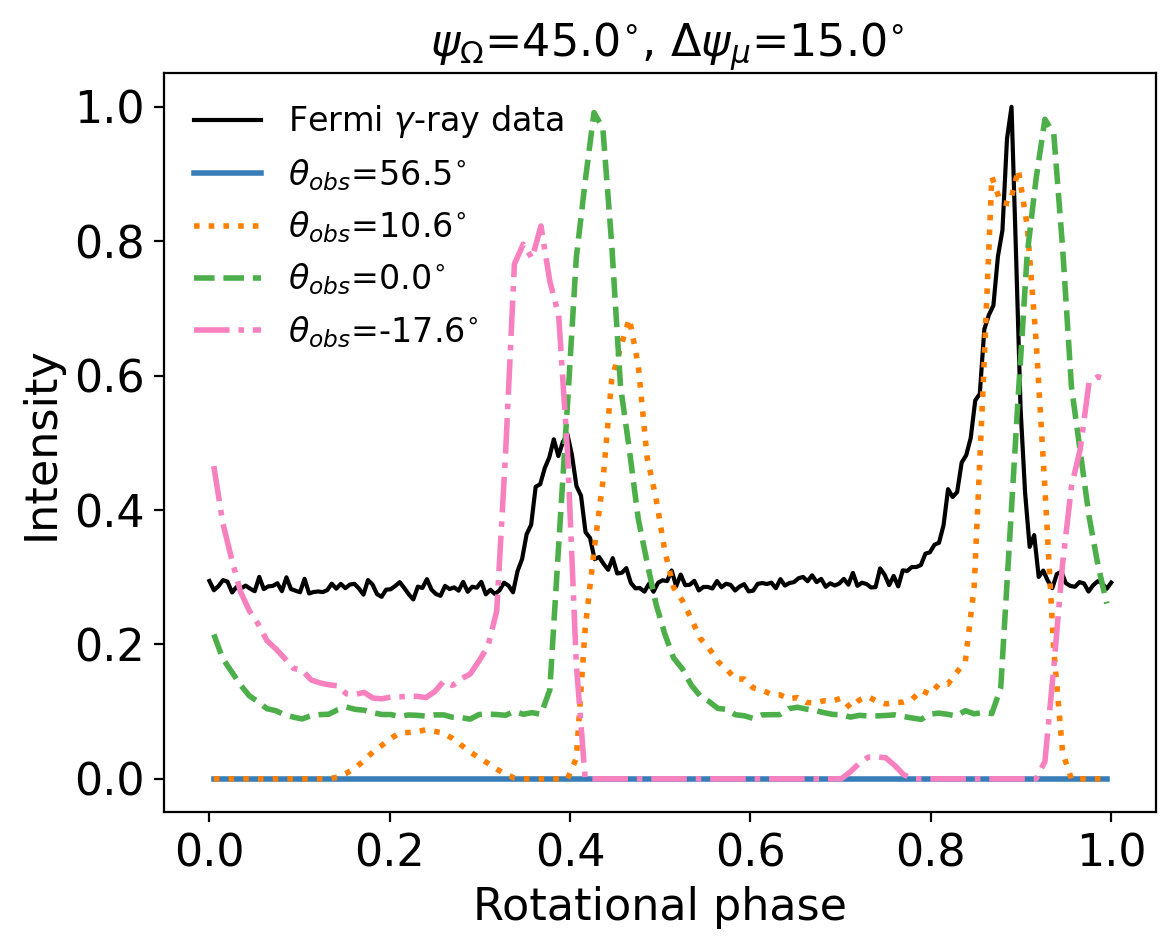}\\
    \includegraphics[width=0.33\textwidth]{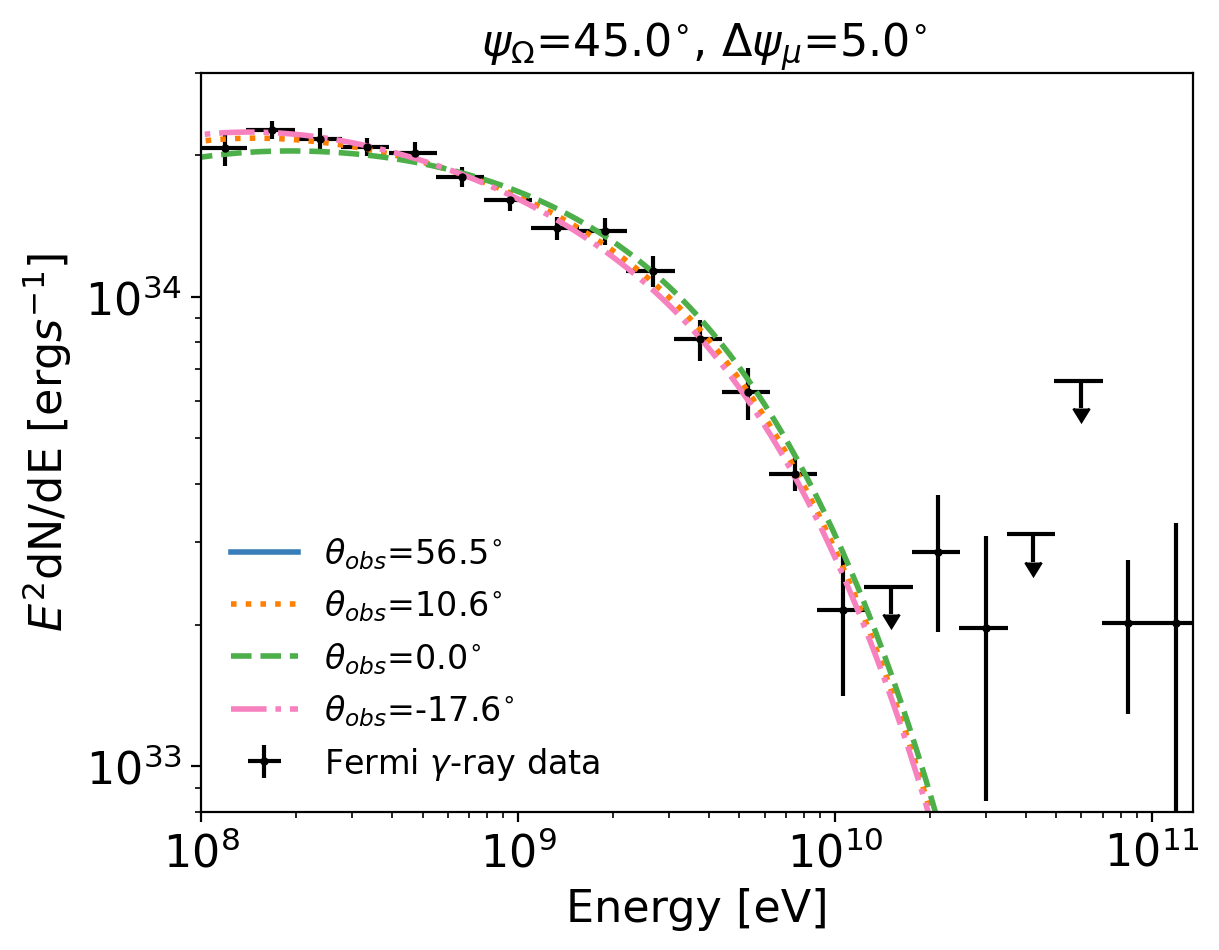}%
    \includegraphics[width=0.33\textwidth]{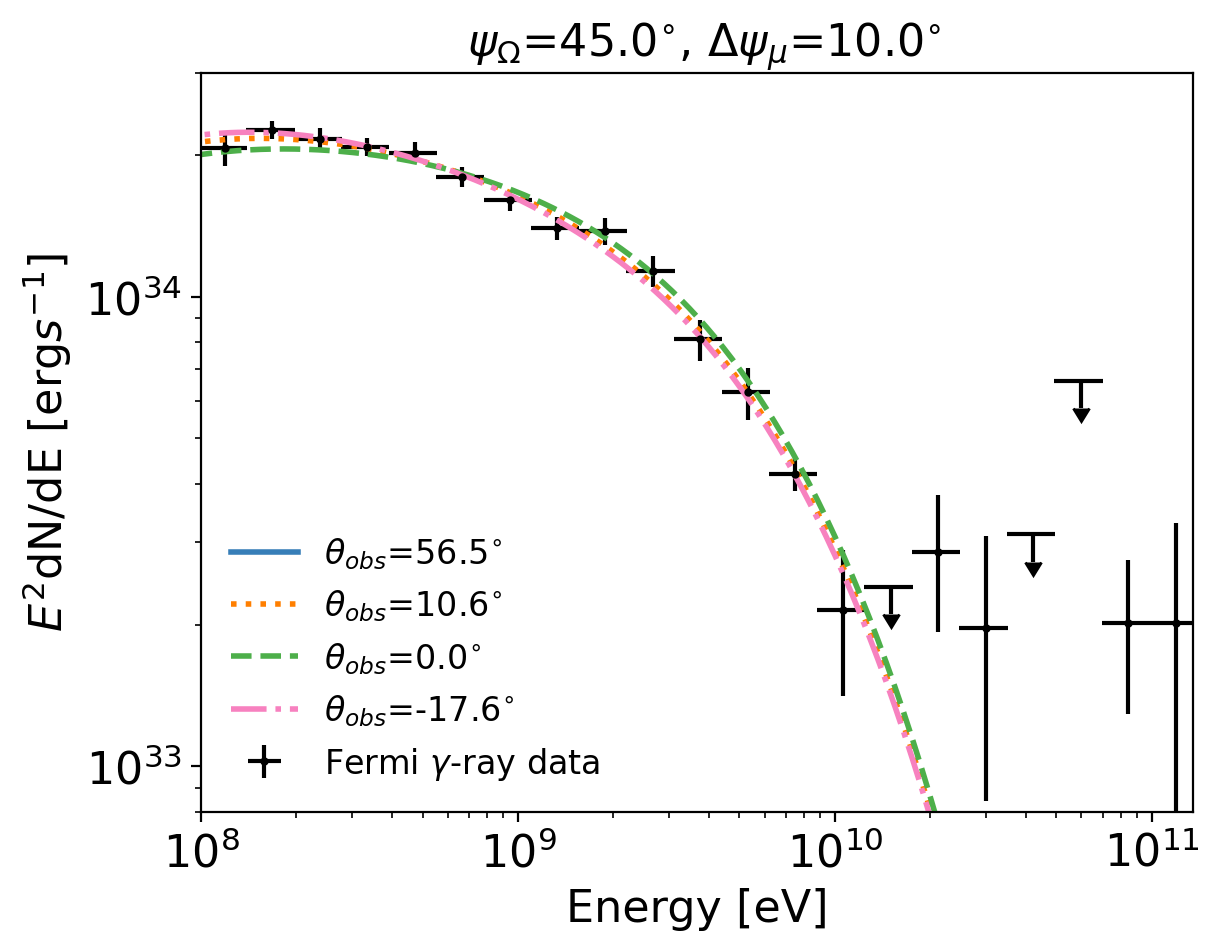}%
    \includegraphics[width=0.33\textwidth]{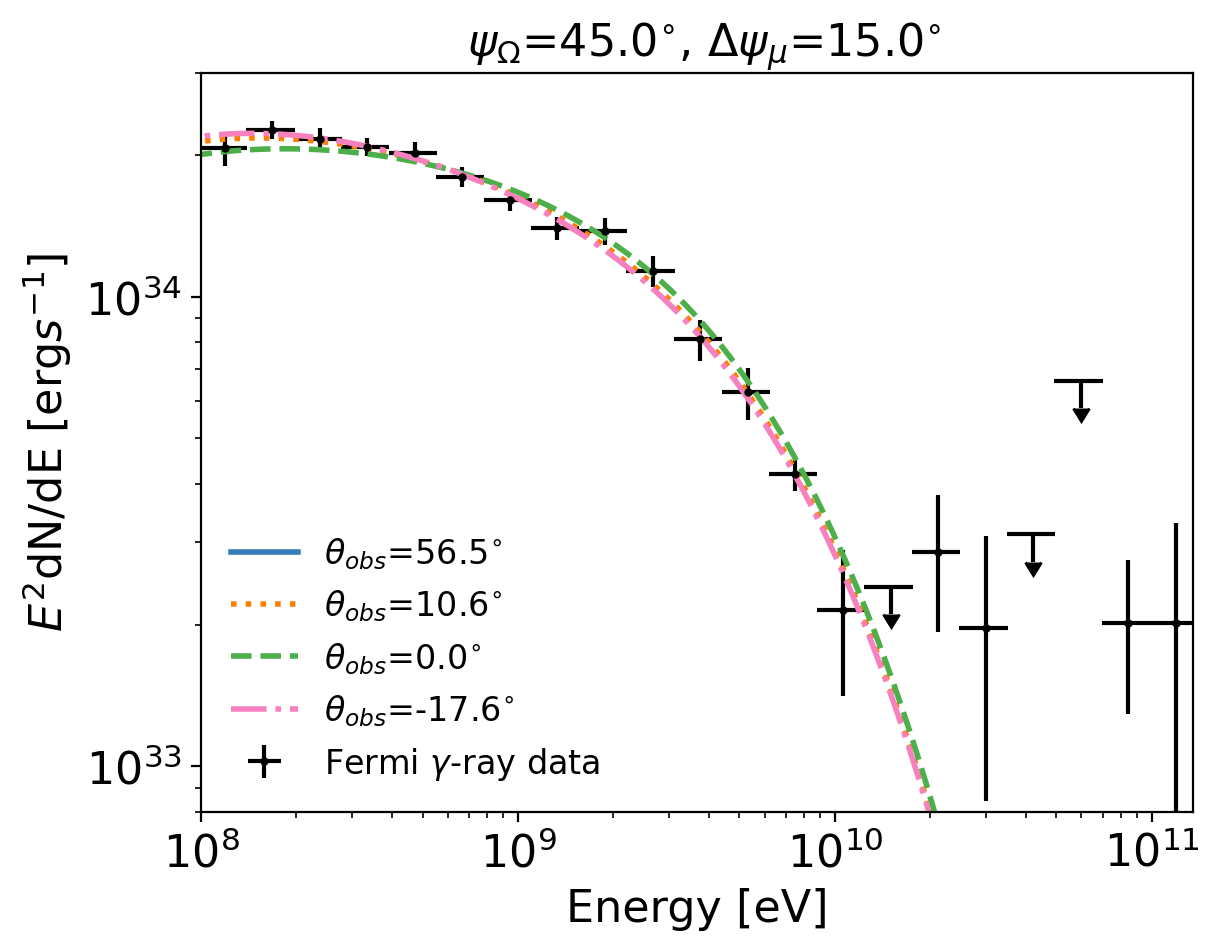}\\
    \caption{Same as Fig. \ref{fig:skymaps_0205_varying_psiomega}, but varying the meridional extent of the region $\Delta\Psi_{\mu}=\{5^\circ,10^\circ,15^\circ\}$ (from left to right), for an inclination angle $\Psi_\Omega=45^ \circ$ and an injection range $\Delta R=0.5~R_{lc}$.}
    \label{fig:skymaps_0205_varying_deltapsimu}
\end{figure*}

Let us start by studying the effect of the geometrical parameters for a given pulsar, focusing on J0205+6449.
The second rows of Figs. \ref{fig:skymaps_0205_varying_psiomega} and \ref{fig:skymaps_0205_varying_deltapsimu} show a selection of skymaps for different geometries, visualized in the first rows (recall that these skymaps, as well as their corresponding light curves, are obtained by integrating $M_E$ in energies, in the $\gamma$-ray band, 100 MeV -- 300 GeV). 
In each figure we can separately see the impact on the emission map of the two free geometrical parameters. 
Second row of Fig. \ref{fig:skymaps_0205_varying_psiomega} shows how the skymap is clearly modified when the inclination angle varies. 
For low inclinations, the emission is concentrated around the equatorial plane, whereas more structures appear for larger inclination angles.
On the other hand, second row of Fig. \ref{fig:skymaps_0205_varying_deltapsimu} shows the broadening of the intensity region when $\Delta\Psi_{\mu}$ increases. 
Looking at Fig. \ref{fig:plot_region_with_angles} we can see how the region bends as a function of the radial distance to the star, although in the innermost part of the region this bending is not very strong. 
In this way, there is a certain radial extent $R$ of the region in which the particles point mostly in the same direction, or in a small angular range. 
Therefore, their emission is concentrated in a band, with a width around $\Delta\Psi_{\mu}$, which is what we see on the skymaps. 
At the same time, in the outer parts of the region, the bending is stronger, meaning that the particles there point to a broader angular range, thus spreading more their emission, as is also seen in the skymaps.

The third rows of Figs. \ref{fig:skymaps_0205_varying_psiomega} and \ref{fig:skymaps_0205_varying_deltapsimu} show examples of several light curves of the same pulsar, for the same particular set of geometrical parameters of its corresponding skymap on the second row, and different viewing angles. 
We observe how the synthetic light curves we generate have general shapes compatible with the observational ones found in the 2PC and 3PC. 
This can be seen by qualitatively comparing the colored lines (synthetic) against the black line (observational from the 3PC). 
Note that we are only showing some particular geometries, not necessarily the best-fitting ones.
Our set shows a wide diversity of light curve features, having one up to several peaks, with different peak separations, and peak widths. 
The goal of this work is indeed to show that our spectral+geometrical model generates realistic light curves, rather than doing quantitative fits (a task left for future work).

Light curves obtained from geometries with large inclination angles present a richer morphology than those coming from lower inclinations. 
All the non-zero light curves in the left panel of the third row of Fig. \ref{fig:skymaps_0205_varying_psiomega} have two peaks, while those in the middle and right panel also show three and four peaks.

We also note that for higher inclination angles the skymap gets more filled, because the whole region is more inclined and the particles in it can span a broader range of emission directions. 
This can be clearly seen with the observer $\theta_{obs} = 56.5^{\circ}$, the one closer to the pole. For $\Psi_{\Omega} = 27^{\circ}$, this observer would not detect anything, while for $\Psi_{\Omega} = 54^{\circ}, 81^{\circ}$, she would.
Increasing the meridional extent does not have such a critical impact on the light curves. 
It modifies the width of the peaks, making them wider, as can be seen in the third row of Fig. \ref{fig:skymaps_0205_varying_deltapsimu}.
It also has the effect of filling the skymaps, to a lower degree than with the increase of the inclination angle, and at the same time increases the intensity of the regions where radiation is collected. 
This latter effect occurs because increasing the meridional extent implies the presence of more lines along the magnetic colatitude. Since the angular increase is rather small (in our case, $2.5^{\circ}$ at each hemisphere when increasing from $\Delta\Psi_{\mu} = 5^{\circ} (10^{\circ})$ to $\Delta\Psi_{\mu} = 10^{\circ} (15^{\circ})$, or $5^{\circ}$ at each hemisphere when increasing from $\Delta\Psi_{\mu} = 5^{\circ}$ to $\Delta\Psi_{\mu} = 15^{\circ}$), those new lines have a very similar direction than those already present with a smaller $\Delta\Psi_{\mu}$. 
In this way the amount of emitting particles (which point in similar directions) increases, and thus more radiation is collected in the sections of the skymap already populated in the smaller $\Delta\Psi_{\mu}$ case, increasing the intensity of these sections.

Finally, we note that some observers, e.g. the one located at $\theta_{obs} = 56.5^{\circ}$ in Fig.~\ref{fig:skymaps_0205_varying_deltapsimu}, would see no emission at all, at any phase.
This  likely  happens in nature too, i.e., there could be high-energy pulsars as powerful as the one taken here as example, J0205+6449, which we do not detect simply because we are not in the line of sight of their emission.
Below we will assess the percentage of observers which actually do detect emission in our skymaps.

\begin{figure}
    \centering
    \includegraphics[width=0.25\textwidth]{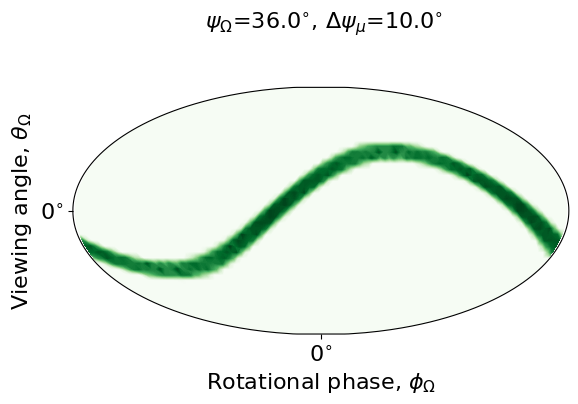}%
    \includegraphics[width=0.25\textwidth]{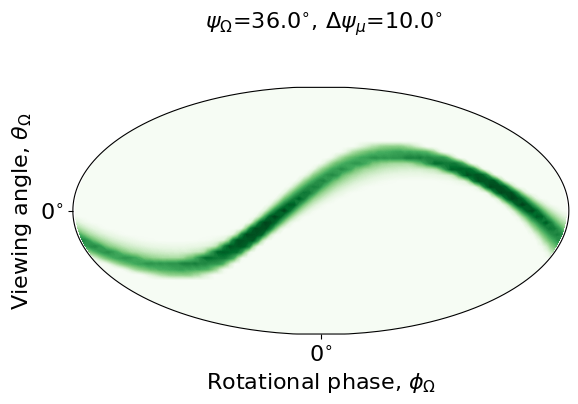}\\
    \includegraphics[width=0.25\textwidth]{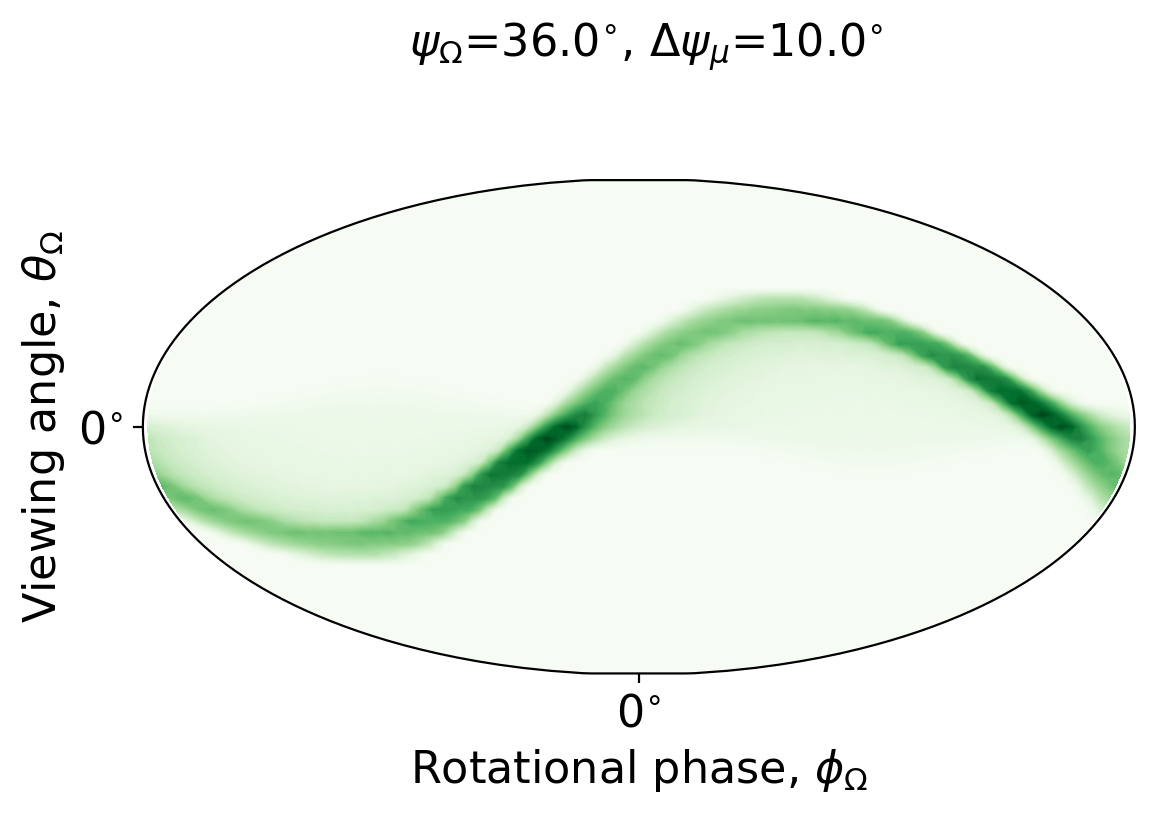}%
    \includegraphics[width=0.25\textwidth]{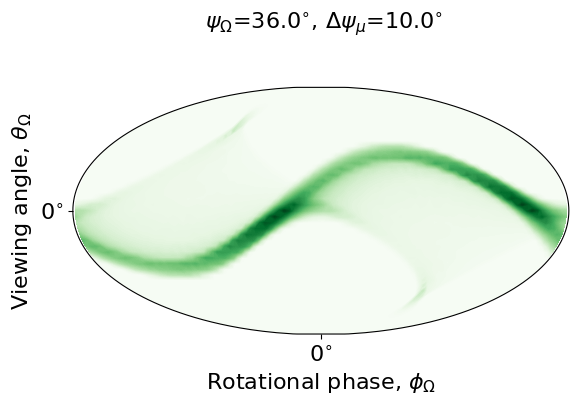}\\
    \caption{Skymaps of J0205+6449 for a particular geometry, $\Psi_{\Omega} = 36^{\circ}$ and $\Delta\Psi_{\mu} = 10^{\circ}$, and different values of $\Delta R$. These skymaps are obtained by integrating $M_E$ in energies, in the $\gamma$-ray band, 100 MeV -- 300 GeV. Top row (from left to right): $\Delta R = 0.0, 0.25$. Bottom  row (from left to right): $\Delta R = 0.50, 0.75$.}
    \label{fig:skymaps_different_deltaR}
\end{figure}

Finally, Fig. \ref{fig:skymaps_different_deltaR} shows the skymaps resulting from different values of $\Delta R = \{ 0, 0.25, 0.50, 0.75\}~ R_{lc}$ (keeping the same radial injection resolution, i.e. having 1, 11, 21 and 31 injection points, respectively), considering one particular geometry, $\Psi_{\Omega} = 36^{\circ}$ and $\Delta\Psi_{\mu} = 10^{\circ}$.
We observe how the increasing of the injection range has the effect of filling more sections of the skymap, due to the modulation of the outer parts of the trajectory (see Fig. \ref{fig:plot_region_with_angles}).
A larger injection range produces a more elongated region, and thus the radiation emitted spans a broader set of directions, as seen in the skymaps, which are more filled for a larger $\Delta R$.
In any case, fixing $\Delta R = 0.5 R_{lc}$ produces region shapes more similar to those obtained with PIC simulations than other values.

\subsection{Effects of the spectral parameters on the skymaps and light curves}\label{effects_different_pulsars_sms_lcs}

\begin{figure}
    \centering
    \includegraphics[width=0.33\textwidth]{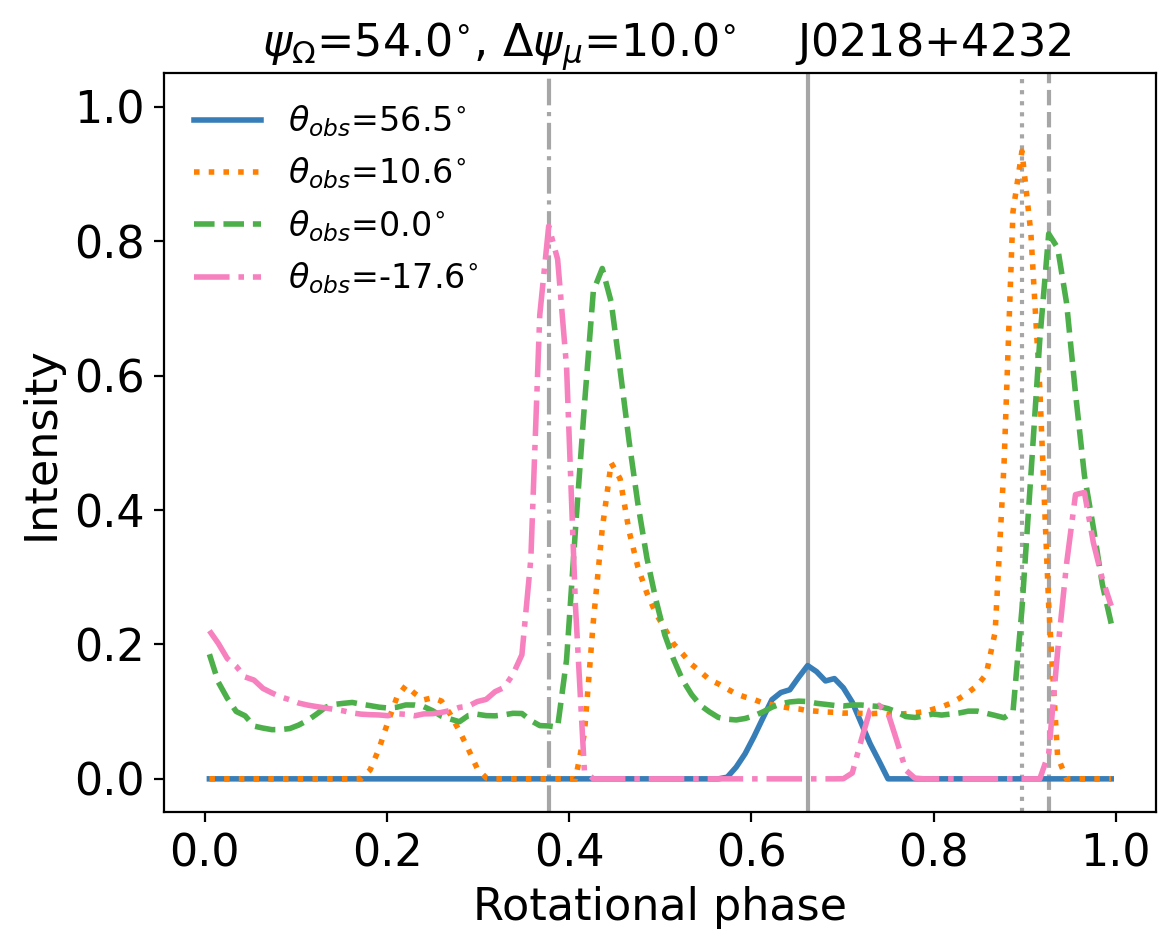}\\
    \includegraphics[width=0.33\textwidth]{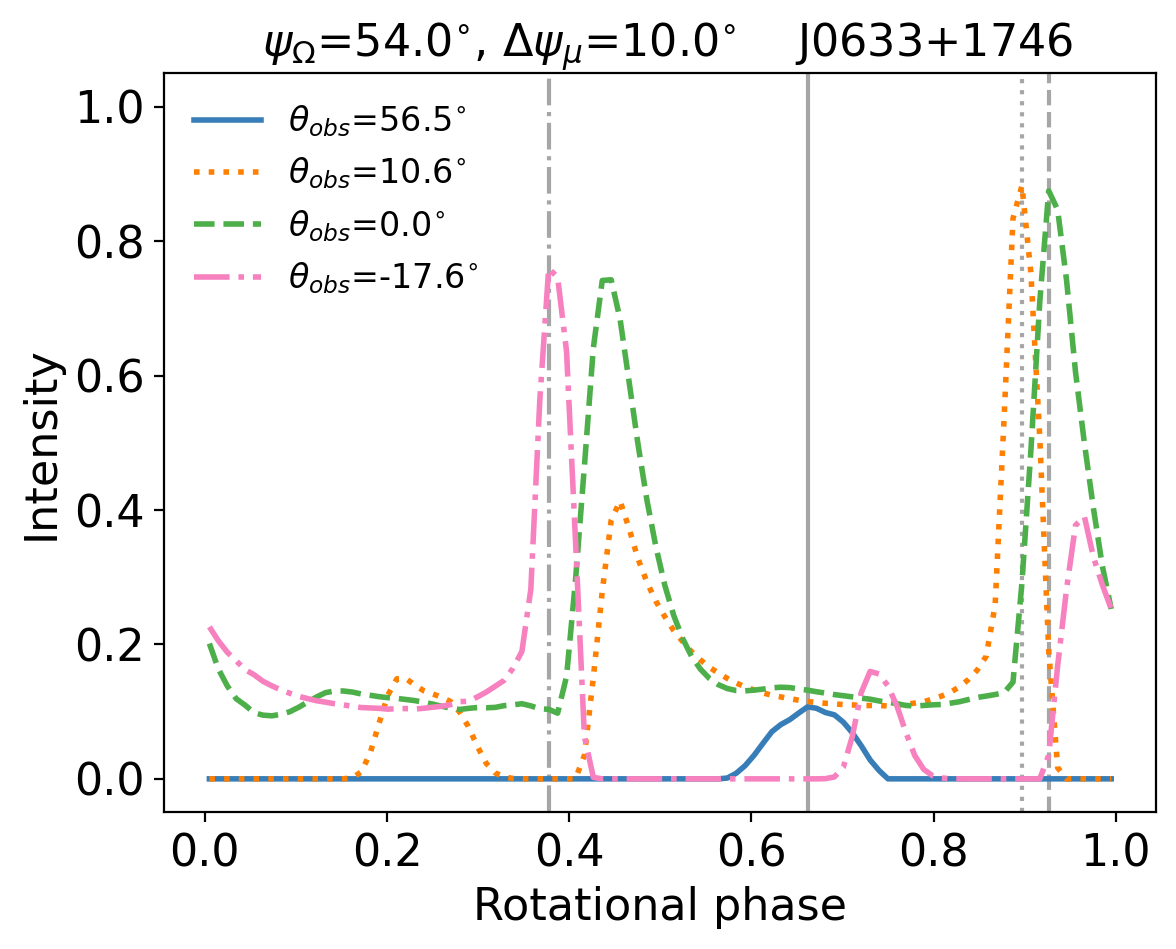}\\
    \includegraphics[width=0.33\textwidth]{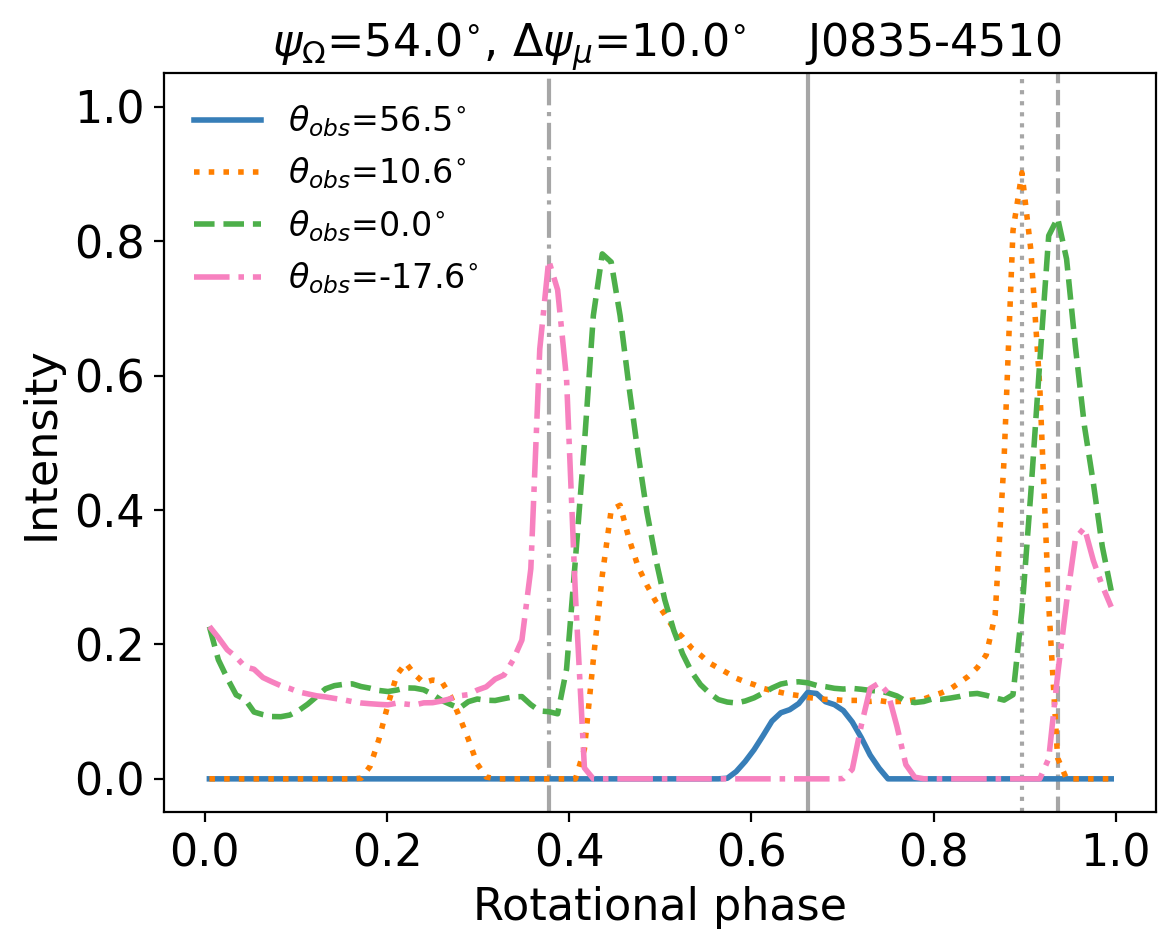}\\
    \caption{Selection of light curves of the particular geometry $\Psi_{\Omega} = 54^{\circ}$ and $\Delta\Psi_{\mu} = 10^{\circ}$, for three pulsars, from top to bottom: J0218+4232, J0633+1746 and J0835-4510. Light curves are obtained by integrating $M_E$ in energies, in the $\gamma$-ray band, 100 MeV -- 300 GeV, and cutting an horizontal slice. Vertical grey lines indicate the phase of the highest-intensity peak of each light curve, which basically don't change between the light curves of different pulsars having the same geometry. }
    \label{fig:lightcurves_different_pulsars}
\end{figure}

So far we have seen the effects of the geometry, i.e. of the geometrical parameters, for a given pulsar. Let us now compare different pulsars, i.e. different spectral properties of the emission, for the same geometry. 
In Fig. \ref{fig:lightcurves_different_pulsars} we show the light curves generated for three of our selected pulsars, for a particular geometry, $\Psi_{\Omega} = 54^{\circ}$ and $\Delta\Psi_{\mu} = 10^{\circ}$, the one corresponding to the central panels of Fig. \ref{fig:skymaps_0205_varying_psiomega}.
They are basically similar, with only small variations on the intensity. Therefore, changing the spectral (and the timing) properties alone produce changes on the skymap which are barely visible, in any case much smaller than the effects produced when the geometry is changed.

This implies that the geometrical parameters are much more important than the spectral ones in determining the light curves.

\subsection{Spectra seen by different observers}

If we integrate azimuthally $M_E(\theta_{obs}, \phi_\Omega, E)$ for a given observer, we obtain the phase-average SED seen by it.
Therefore, once we build a map, we can get the spectra all the observers would detect.

Bottom rows of Figs. \ref{fig:skymaps_0205_varying_psiomega} and \ref{fig:skymaps_0205_varying_deltapsimu} show the synthetic spectra seen by selected observers.
Even though differences are not extreme, they have different shapes and resemble the data differently.
The reason for these differences is the non-trivial geometry of the emission region.
Its modulated shape and the fact of having injection points all along the region, has the effect of an observer receiving radiation from particles located at different radial distances from the star as well as at different positions from its injection point into the region.
This implies different values of the kinetic quantities, e.g. Lorentz factor $\Gamma$ and pitch angle $\alpha$, of these particles, which therefore emit different spectra. 

In the light of these results, it is possible to appoint an observer, or at least a group of observers, as preferred, i.e. to give a preferred value or a range of values, for instance, of the possible viewing angle(s), for the particular pulsar studied, by how well the spectra of the observers adjust to the observational data of this pulsar.

We can quantify the degree of resemblance of each spectra by computing a reduced $\chi^2$,  $\overline{\chi^2}$.
Doing so for each observer, we obtain values (evenly distributed) between $\sim 0.8 - 1.7$ (with 18 data points), meaning that the difference among spectra is statistically significant.
In the same way as we did in \cite{sc_emitting_regions_2022}, following the prescriptions of \cite{avni_factor}, we can define a 1 $\sigma$ confidence interval around the smallest $\overline{\chi^2}$ found.
Taking into account that  $\overline{\chi^2}$ values below $1$ can be regarded as statistically equal, for the corresponding degrees of freedom we get a $\overline{\chi^2}_{limit}$ of $1.19$.
Thus, most of the observers have $\overline{\chi^2}$ values outside the 1 $\sigma$ confidence interval of the observer with the lowest $\overline{\chi^2}$.
Still, for different observers, the difference in their spectra is much smaller than the difference in their light curves, as shows the comparison of the third and fourth rows of Figs. \ref{fig:skymaps_0205_varying_psiomega} and \ref{fig:skymaps_0205_varying_deltapsimu}. 

Also, we foresee even larger differences in spectra of different observers when extending the spectra to lower energies, due to an increased dominance of synchrotron radiation.
This will be explored in the future.

\section{Light curves features and comparison with observations}
\label{light_curves_features}

Out of the total number of synthetic light curves generated for a given pulsar as a result of the application of our model, the percentage of detected light curves, i.e. of observers that do see a non-zero intensity, is typically $60\%$. The remaining are non-detections, i.e. observers whose viewing angles do not cut through the bulk of the emitted radiation, corresponding to pulsars we do not detect due to geometrical reasons. This result may have an impact in population synthesis studies, since it gives an estimate of how many pulsars can indeed be detected given sufficient sensitivity out of the total real population.

We now consider the distribution of morphological features of the detected light curves:
\begin{itemize}
    \item Number (or percentage) of light curves with $n$ peaks 
    
    \item Flux ratio between two peaks (P1/P2, where  P1 is taken as the highest) 
    
    \item Phase separation between two peaks
    
    \item Width of the peaks 
\end{itemize}

For the first one, we compute the
percentages over detected light curves (i.e. discounting from
the total number of light curves generated (1530) the non-detections).
For each percentage we associate an uncertainty of $3 \%$. 
Such a value is the typical maximum variation that includes the uncertainty due to both the resolution of the region size, and the algorithm that defines the peaks (in particular, changing the threshold cut between 5\% to 1\%, see Appendix \ref{app:peak_recognition_algorithm} for details).

\begin{figure}
    \centering
    \includegraphics[width=0.45\textwidth]{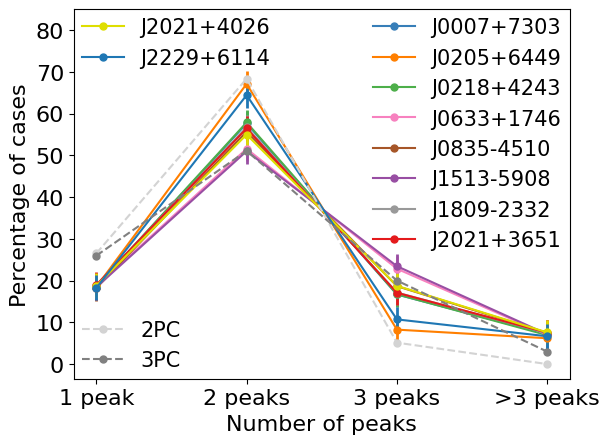}
    \caption{Percentages of synthetic light curves with $n$ peaks for different pulsars. Grey lines show the observed distribution (from all pulsars) in the 2PC (117) and the 3PC (236). Error bars of $3\%$ indicate the estimated uncertainties as explained in the text.}
    \label{fig:statistics_peaks_different_pulsars}
\end{figure}

\begin{figure*}
        \centering
        \includegraphics[width=0.33\textwidth]{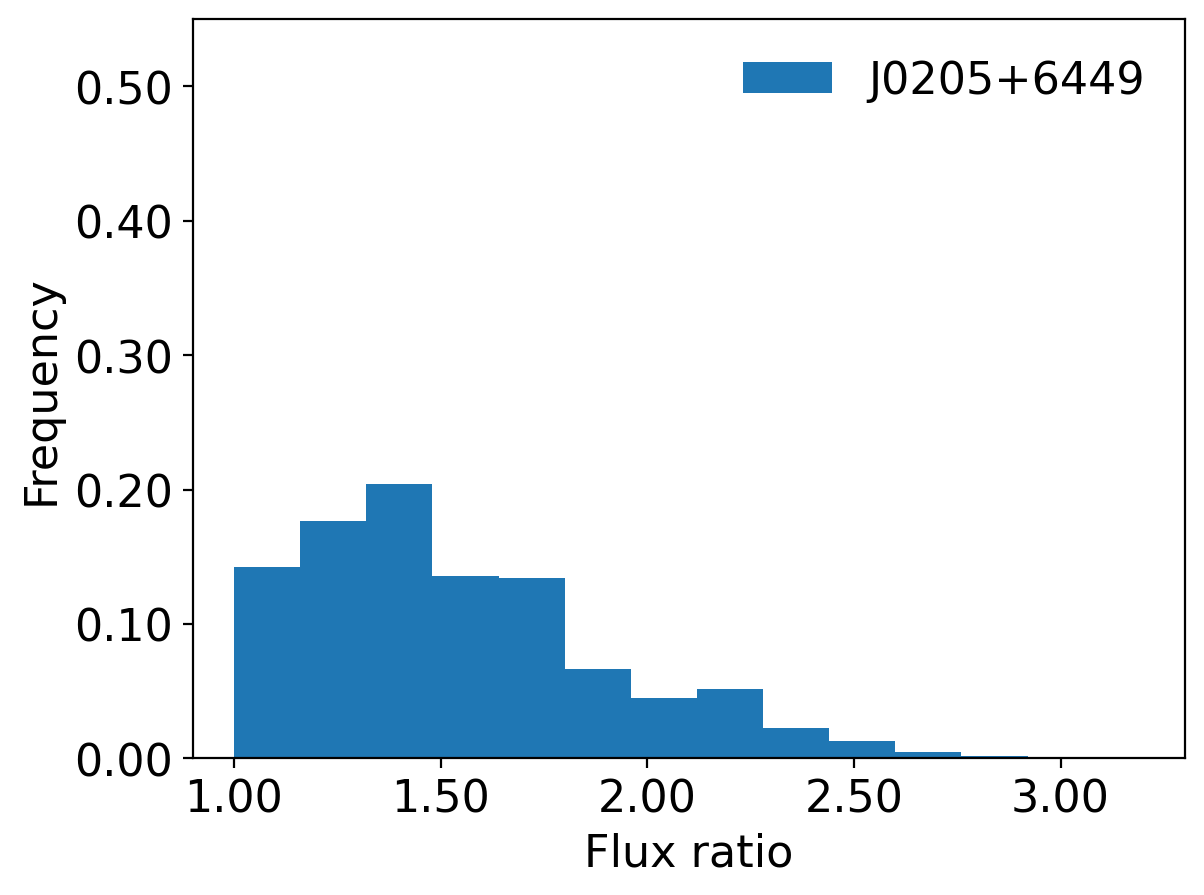}%
        \includegraphics[width=0.33\textwidth]{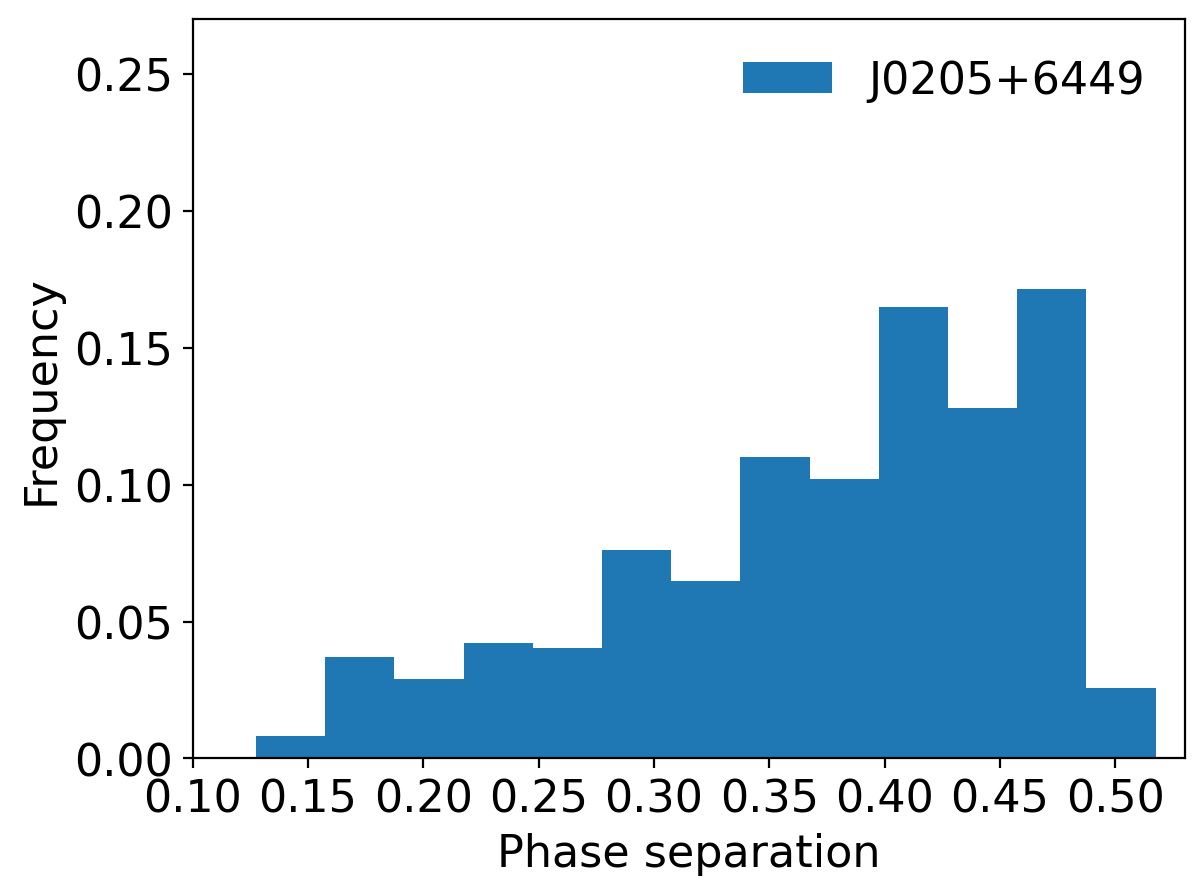}%
        \includegraphics[width=0.33\textwidth]{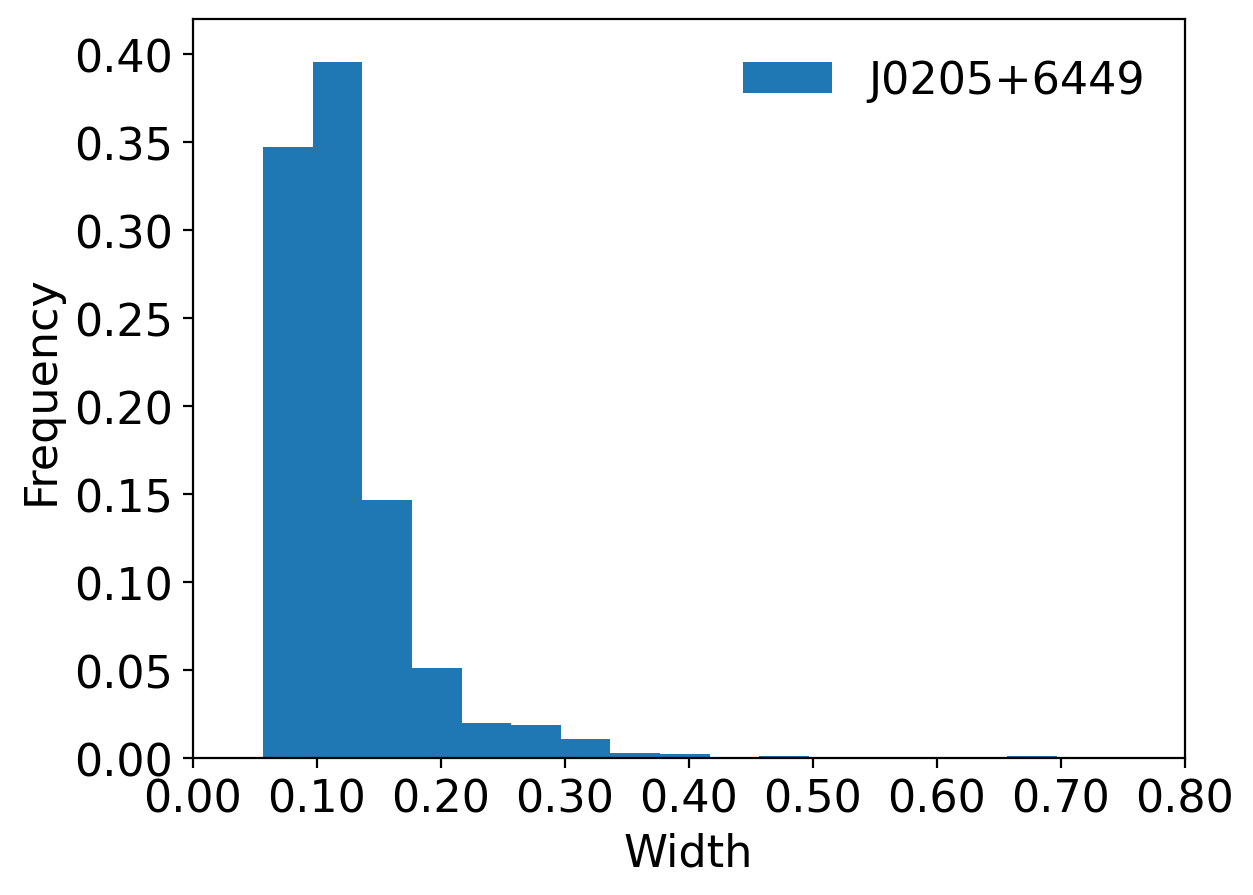}\\
        \includegraphics[width=0.33\textwidth]{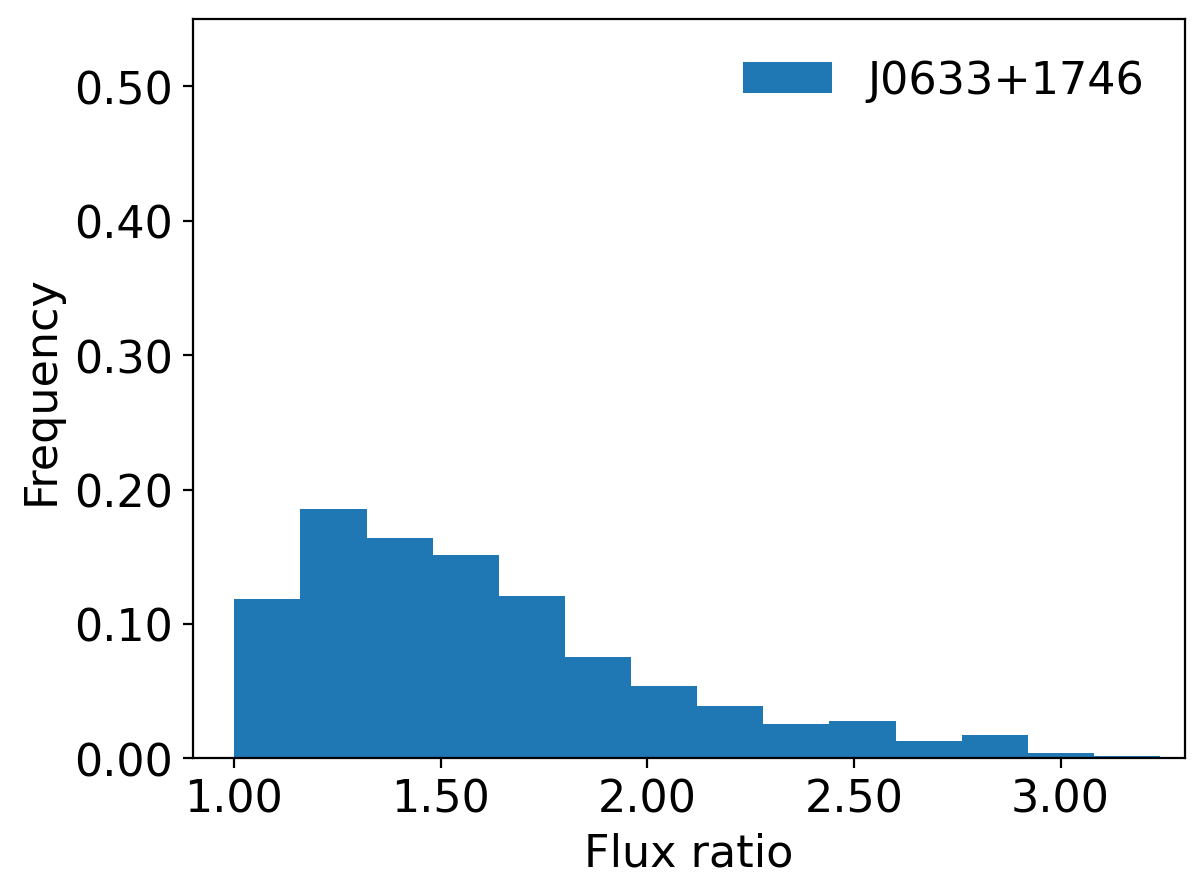}%
        \includegraphics[width=0.33\textwidth]{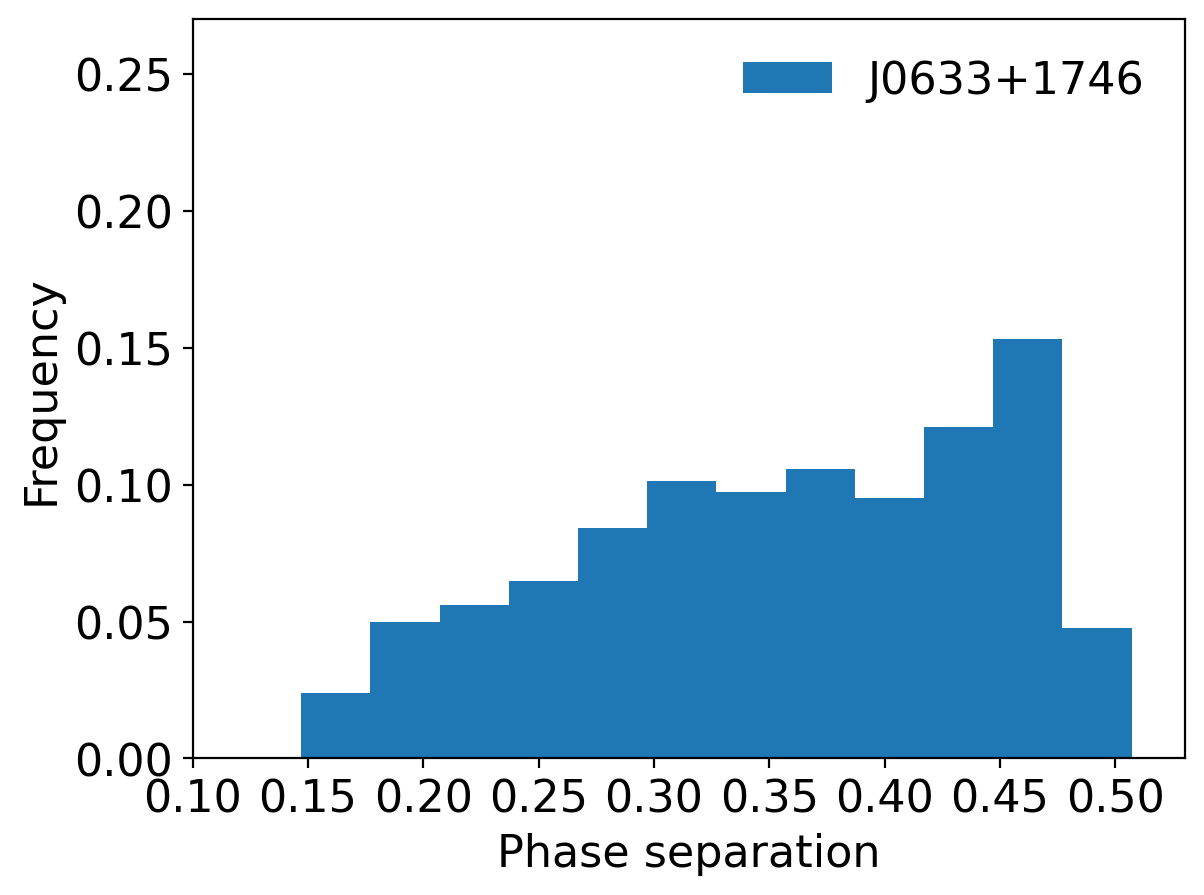}%
        \includegraphics[width=0.33\textwidth]{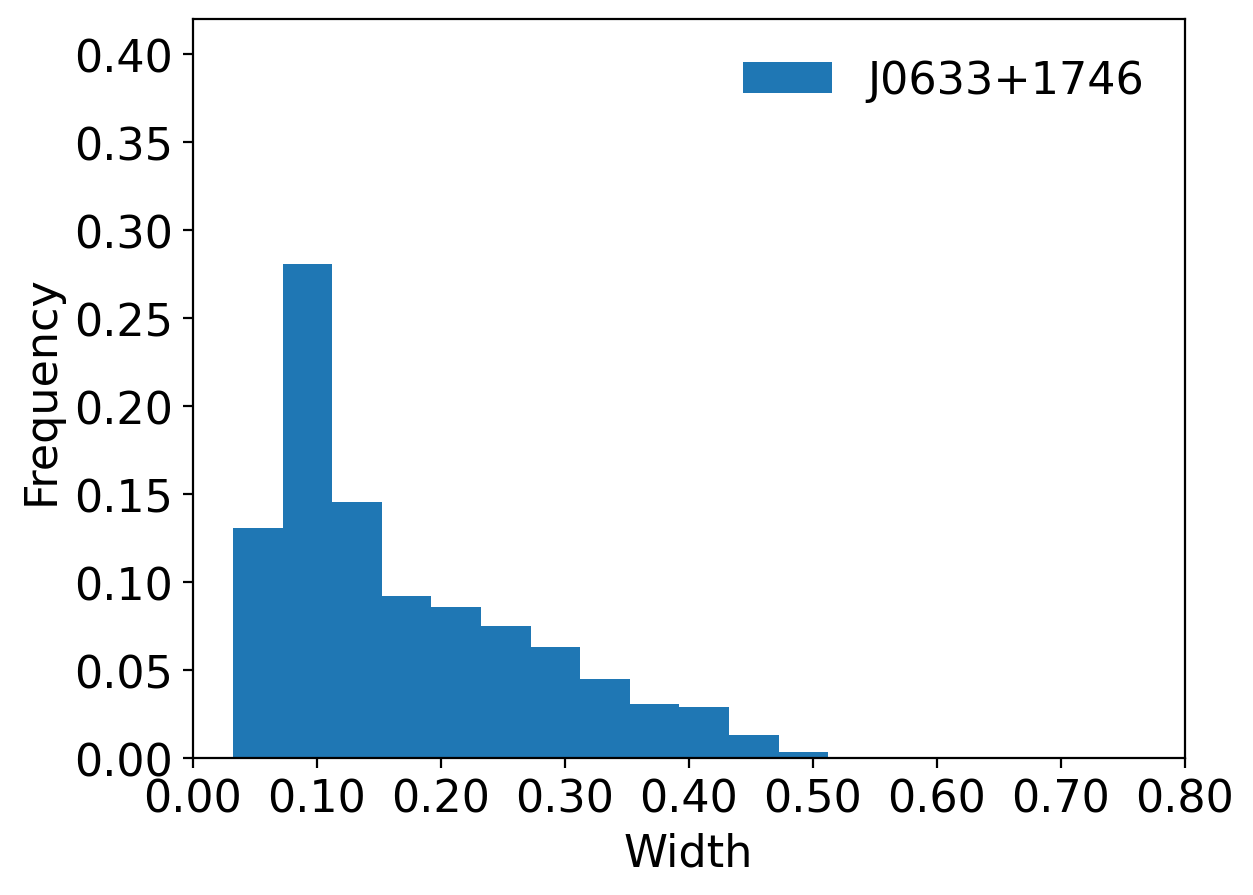}\\
        \includegraphics[width=0.33\textwidth]{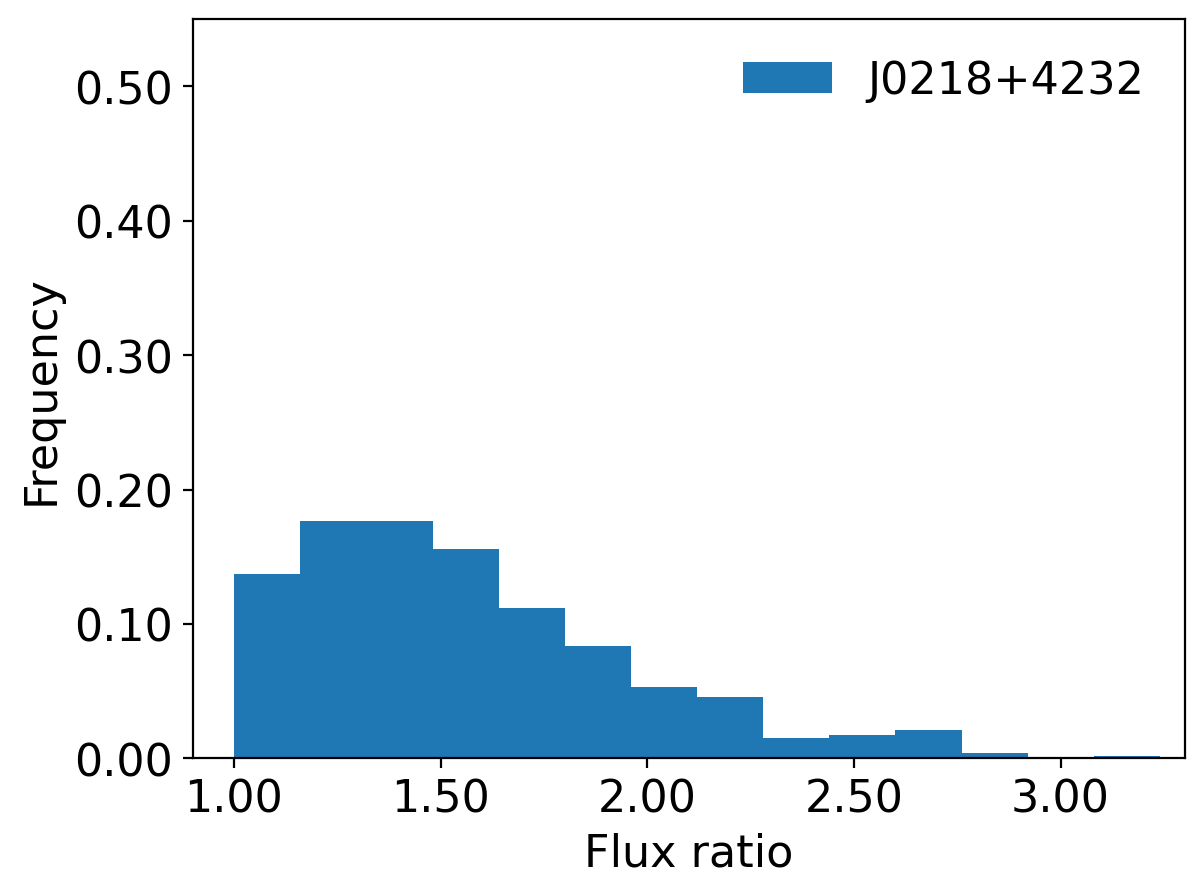}%
        \includegraphics[width=0.33\textwidth]{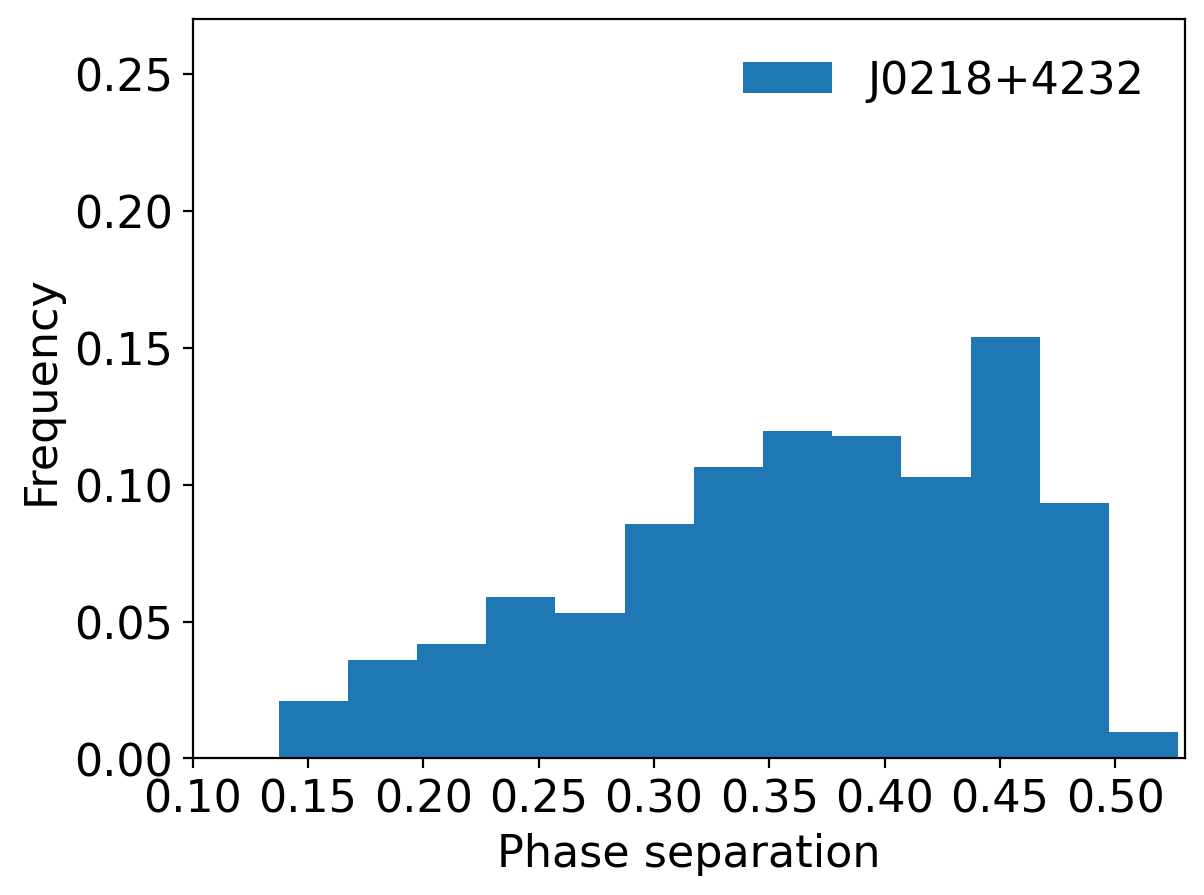}%
        \includegraphics[width=0.33\textwidth]{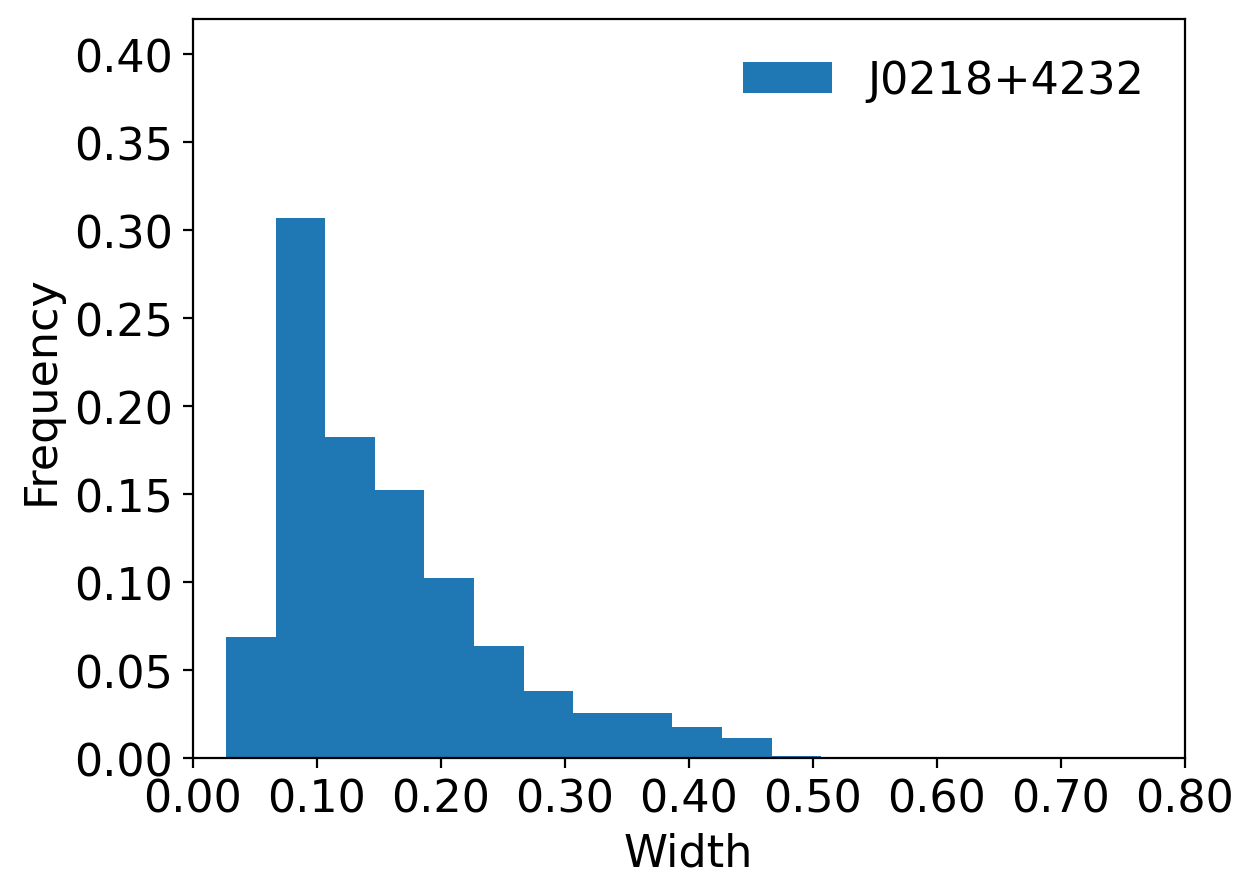}\\
        \includegraphics[width=0.33\textwidth]{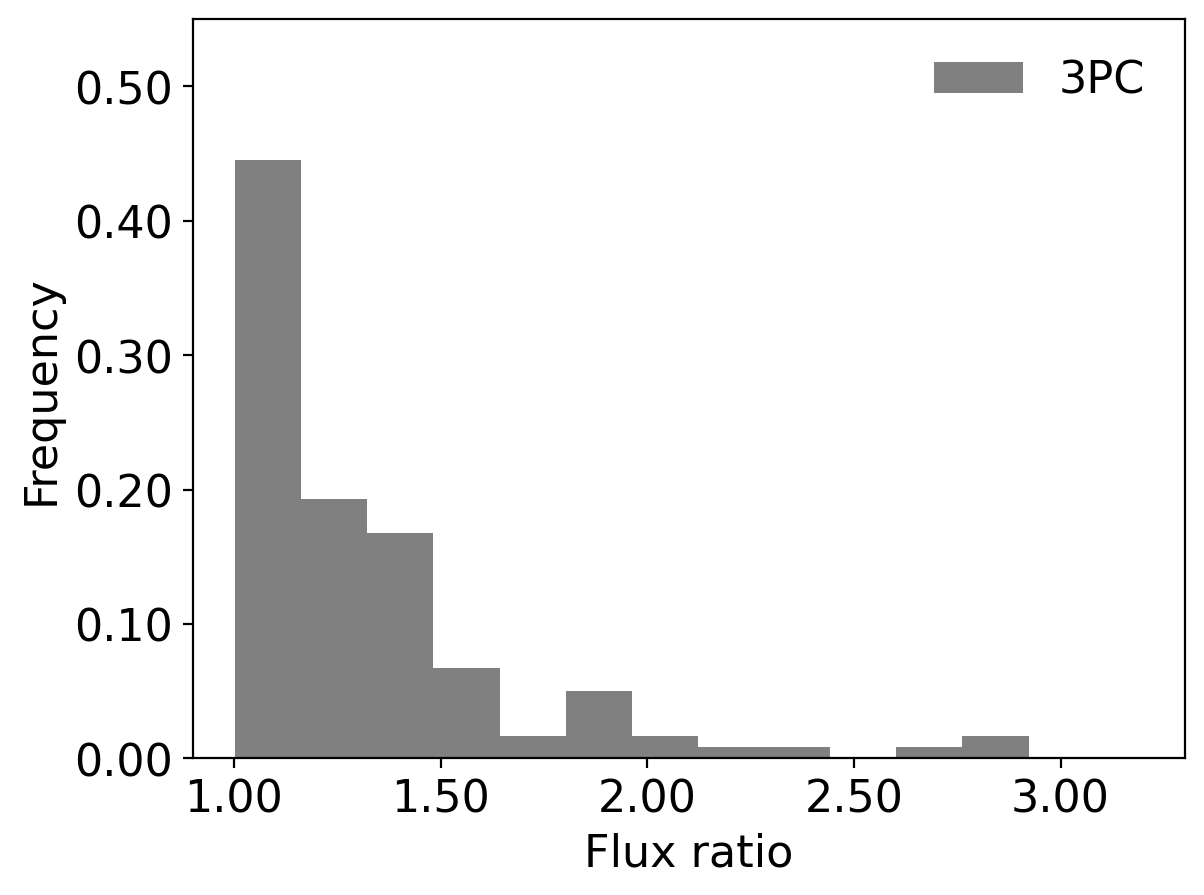}%
        \includegraphics[width=0.33\textwidth]{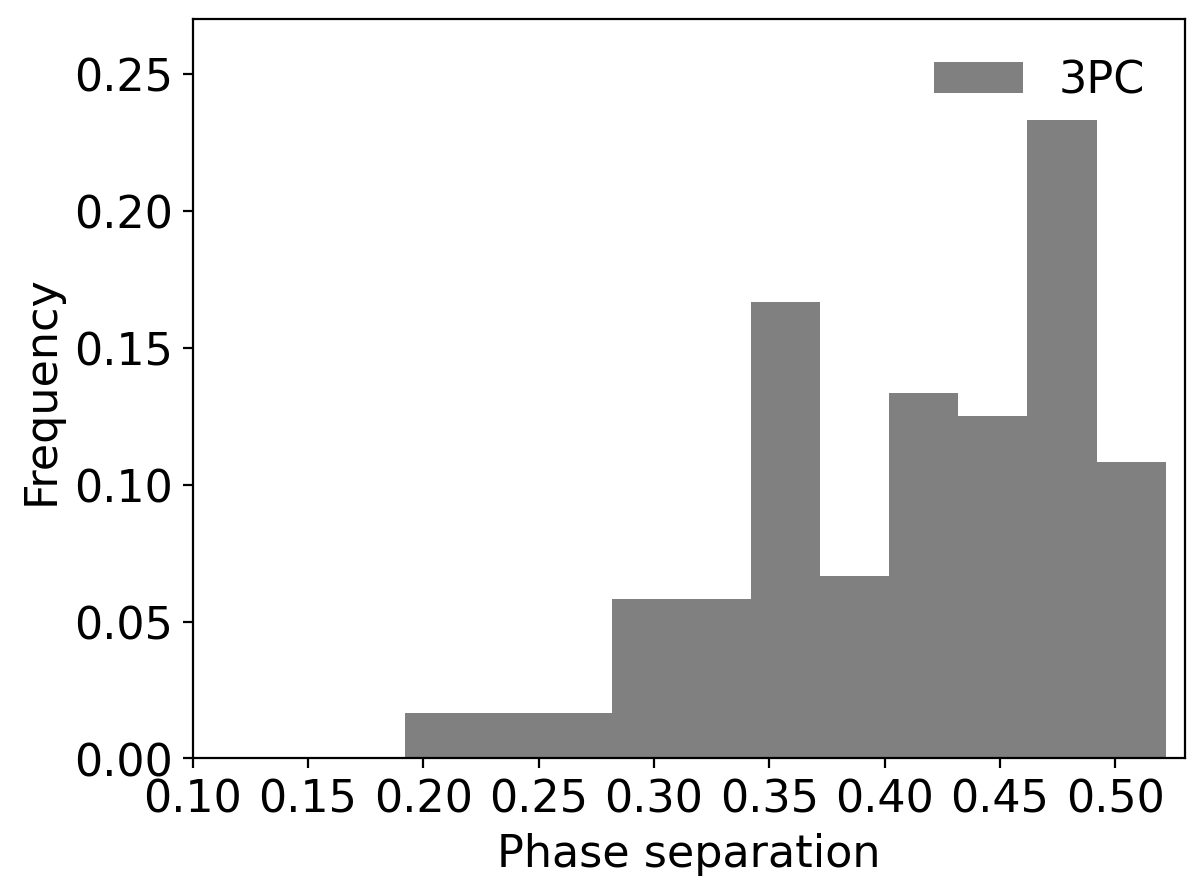}%
        \includegraphics[width=0.33\textwidth]{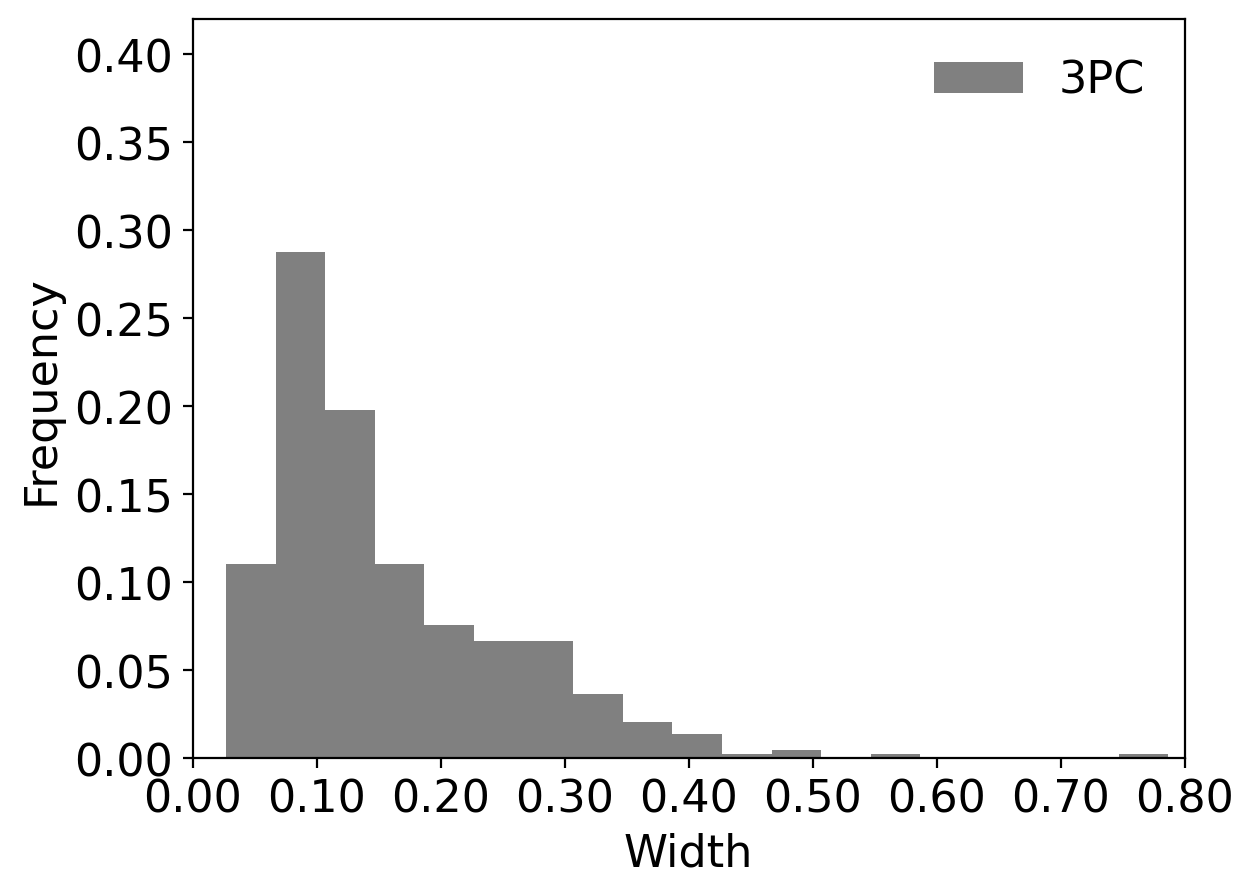}\\
    \caption{Normalized frequency histograms of flux ratio (left column), phase separation (middle column) and width of the main peak (right column), coming from the synthetic light curves over geometry variations for three pulsars (first three rows, respectively J0205+6449, J0633+1746 and J0218+4232), and from the observed pulsar 3PC sample (last row). For the latter, we have used the same peak recognition algorithm, except for the width which has been computed manually.} 
    \label{fig:histograms_indicators_different_pulsars}
\end{figure*}

The second and third features (flux ratio and phase separation) apply to the subset of light curves with just two peaks, and the fourth one (width) is calculated for each one of the peaks in each light curve.
We construct histograms for each of these three features. Fig. \ref{fig:statistics_peaks_different_pulsars} shows the percentages of light curves with $n$ peaks
for the selected pulsars.
Fig. \ref{fig:histograms_indicators_different_pulsars} shows examples of the histograms for the other three features.

In Fig. \ref{fig:statistics_peaks_different_pulsars} we can see that most of the detected light curves possess two peaks, while few have more than 3 peaks. 
The agreement with the global result of the 2PC/3PC is impressive. We comment more on this below.

Fig. \ref{fig:histograms_indicators_different_pulsars} shows clear trends on the distributions of the flux ratio and width of the peaks. 
The former peaks at $\sim 1.3$ and decreases up to $\sim 3$, indicating the rather small difference in the intensity of each peak in two-peaked light curves. 
The widths distribution also peaks at low values, with the majority of the peaks having a width of $0.05-0.30$ in phase and almost no peaks with a width larger than $0.5$.
In the distributions of the phase separation between peaks such a behavior is less pronounced. Still, we do see a preference
for larger separations than not.

\subsection{Comparing the distribution of features of different pulsars} 

Figs. \ref{fig:statistics_peaks_different_pulsars} and \ref{fig:histograms_indicators_different_pulsars} show that the statistics of the different features of the light curves, while being different for different pulsars, follow the same trend.
This fact agrees with what we have seen in subsection \ref{effects_different_pulsars_sms_lcs}, where the skymaps and light curves are very similar among different pulsars.
This reinforces the idea that the geometry plays a major role in shaping the light curves, being much relevant than the spectral parameters.

Fig. \ref{fig:statistics_peaks_different_pulsars} shows a line corresponding to the percentages of light curves with $n$ peaks in the 2PC and 3PC. 
We observe how it follows a very similar behavior to the lines of particular pulsars. The similarity becomes even larger with the 3PC, which has brought better and more $\gamma$-ray light curves than those in the 2PC. 
The better signal-to-noise ratio of the new catalog has brought the appearance of new peaks which in the previous one were overtaken by noise, resulting in the presence of more light curves with 3 peaks or more. It has also provoked the detection of peaks in light curves that were previously too noisy.

This implies that a set of synthetic light curves produced for a fixed set of best-fit spectral parameters but varying the geometry (inclination, observer line of sight) is also representative of the global set of all observed pulsars, for which the values of such angles are also randomly distributed.

\begin{figure}
    \centering
    \includegraphics[width=0.45\textwidth]{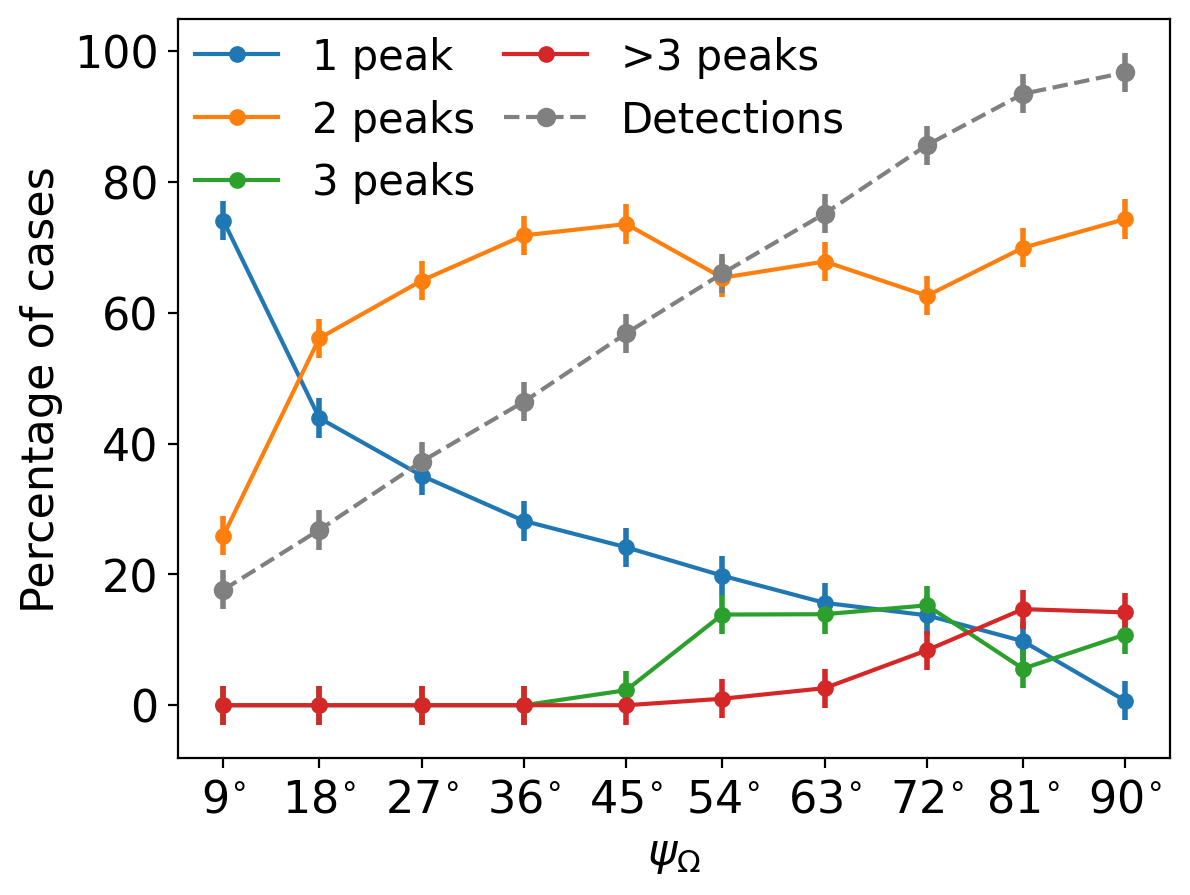}\\
    \caption{Percentages of synthetic light curves (calculated over the observers seeing some signal) with $n$ peaks as a function of the inclination angle $\Psi_{\Omega}$ for J0205+6449. Gray dashed lines represent the percentage of observers detecting signal, calculated over the total set of light curves generated.
    Error bars represent the estimated statistical $3\%$ absolute error on the percentages, as explained in the text.}
\label{fig:statistics_peaks_different_psiomegas}
\end{figure}

\subsection{Impact of the geometrical parameters on the distribution of features}

Having a large sample of synthetic curves, we can look for correlations between the geometrical parameters and the distribution of the values representing the light curve features.
Here we show how the percentages of light curves with $n$ peaks differ for different values of $\Psi_{\Omega}$ and $\Delta\Psi_{\mu}$.
Fig. \ref{fig:statistics_peaks_different_psiomegas} show interesting trends, as is the decrease (increase) of the number of light curves with one (three or more than three) peak(s) when increasing the inclination angle. This relates with what we have seen in Figs. \ref{fig:skymaps_0205_varying_psiomega} and \ref{fig:skymaps_0205_varying_deltapsimu}, where a higher inclination angle implied the appearance of more structure in the skymaps and thus of light curves with more features and peaks.
Very similar trends, with different absolute values are observed for the other pulsars.

Taking into account the trends shown, together with the fact that all pulsars show the same behavior and the statement in the previous subsection that the set of light curves created for a particular pulsar can be considered as equivalent to the population of observed light curves, this opens the possibility of constraining the  distribution of the inclination angle in the global pulsar population.

It is also interesting to check how the number of detected and non-detected light curves vary as a function of the parameters. 
Fig. \ref{fig:statistics_peaks_different_psiomegas} also shows how the number of detected light curves, i.e. observers which would detect a non-zero intensity, increases with the inclination angle. The percentage of detected light curves goes from less than $20\%$ to more than $90\%$ as the inclination increases. This shows the possibility that an observational bias could exist towards seeing orthogonal rotators, which can be relevant for population synthesis models or studies of the time evolution of the inclination angle \citep{Philippov14_time_evolution_inclination_angle}.
Again, this is expected from the skymaps on Figs. \ref{fig:skymaps_0205_varying_psiomega} and \ref{fig:skymaps_0205_varying_deltapsimu}, since more sections of the skymap are filled with radiation when increasing the inclination angle.

In the case of the meridional extent, almost no variation is observed. 
The reason is that the increase of $\Delta\Psi_{\mu}$ does not significantly populate new sections of the skymap as the increase of $\Psi_{\mu}$ does, but instead increases the intensity of the regions already populated, as seen in subsection \ref{effects_different_geometries_sms_lcs}.


\section{Conclusions}
\label{conclusions}

In this paper we have significantly enhanced the geometrical model for the pulsar magnetospheric emission regions presented in \cite{light_curves_2019}.  
This model allows to build emission regions with arbitrary shapes and placed in arbitrary locations in a pulsar's magnetosphere,  ultimately depending on two free geometrical parameters, the inclination angle $\Psi_{\Omega}$ and the meridional extent of the region $\Delta\Psi_{\mu}$.
This flexibility allows us to produce very different emission regions and test their ability to produce realistic light curves.

For a given set of geometrical parameters, we build emission maps, or skymaps, in which the azimuthal direction represents the rotational phase of the pulsar and the meridional direction $\theta_{obs}$ represents the viewing angle of a given observer.
From the skymaps we can extract the light curve (and spectrum) a given observer would detect.

The enhancements of the formalism presented comes in two main flavors. 
On the one hand, we have implemented a purely geometrical improvement, allowing a non-zero torsion in the rotating magnetic field lines (seen from an inertial lab frame) and the construction of a realistic shape of the accelerating region (defined as the zone where the trajectories of the particles are located). 
This shape is achieved by having the magnetic colatitude $\Psi_{\mu}$ as a function of the magnetic longitude $\xi_{\mu}$, as well as making this colatitude dependent on the radial distance $R$ and the inclination angle $\Psi_{\Omega}$.
This permits to have a shape of the emission region in qualitative agreement with the magnetospheric topology found by PIC and force-free simulations.
On the other hand, 
our formalism includes the
implementation of the full synchro-curvature radiation mechanism on top of the geometrical model. 
While in \cite{light_curves_2019} we focused only on the geometrical structure (introducing the Frenet-Serret methodology to describe magnetic field lines), here we consider an emission mechanism together with this geometry, describing the radiation generated by the particles that travel through the emission region. 
Synchro-curvature emission is thus intertwined in the geometrical model itself, and we are able to extract light curves from synchro-curvature emission maps, not just from geometrical maps, which are overall consistent with the spectral, phase-averaged prediction.

With this formalism, we have been able to build realistic skymaps, from which we have obtained $\gamma$-ray light curves, for a selected sample of high-energy pulsars. 
The light curves generated resemble reasonably well the observational ones measured by \emph{Fermi}-LAT.
All the features that are seen in observational light curves, such as different number of peaks, diverse morphology, a variety of peak widths, phase separations and flux ratios, are also present in our synthetic sample.

The skymaps and light curves generated for a set of different pulsars possessing good observational data are very similar, with no large observable differences. 
At the same time, these are greatly modified when the geometrical parameters, both the inclination angle $\psi_{\Omega}$ and the meridional extent of the region $\Delta\psi_{\mu}$, change.

In addition, we have computed global features of our light curves sample, which helps us in qualitatively study the synthetic light curves generated.
We have found that, in agreement with the global results of the 2PC/3PC a majority of light curves have two peaks, being a minority those that possess more than three peaks.
The percentage of detected light curves is a $60 \%$ of the total sample, giving an estimate of the pulsars that would not be detected in the gamma-ray regime simply due to geometrical reasons.
At the same time, the distributions of flux ratios and widths of the peaks show the dominant presence of small values of these two features in the sample of synthetic light curves, peaking at small values and decreasing steeply for larger ones. 
On the other hand, the distributions of phase separations show a preference for large separations.
Different pulsars show similar trends on these features when all the possible light curves are considered, pointing to the consistency of the geometrical model as well 
as to the likely similarity among their magnetospheres.
This is impressive when considering that our sample contains very different pulsars, both in energetics and slow down rate, and include normal and millisecond systems.

The fact that considering different pulsars, i.e. to consider different timing and spectral parameters, has such a small impact in the skymaps, implies that the geometry widely dominates over the spectra in shaping the light curves.
Therefore, considering a sample of light curves generated for the same pulsar, but for different geometries and observers, is substantially equivalent to exploring the population of all observational light curves, since each pulsar has its own (unknown) geometry.

Another interesting outcome is that a direct comparison of our synthetic sample with the sample of observational gamma-ray light curves could constraint the geometry.
The trends seen on the percentage of the number of peaks of the light curves as a function of the inclination angle, together with the overall agreement between the synthetic light curves sample for particular pulsars and the global features observed from the pulsar population as a whole, opens the door to the possibility of inferring the global population distribution of the inclination angle of high-energy pulsars in our Galaxy.
We have also observed the drastic increase in the percentage of detected light curves as the inclination angle increases, going from less than $20 \%$  in an almost aligned rotator to more than $90 \%$ in an orthogonal rotator. 
This uncovers a bias bound to be important for population synthesis studies.

In addition, we have shown that different observers see different $\gamma$-ray spectra due to the non-trivial shape of the emission region.
This allows to distinguish observers by the resemblance with observational data of the spectra they see, and at the same time opens the door to the possibility of giving possible value(s) of the viewing angle of the pulsar being studied.
The observed difference in spectra are enlarged when the full spectra is considered. In particular, if the non-thermal X-ray emission is considered together with the $\gamma$-ray radiation, as we did in our previous spectral-only studies \citep{diego_solo, systematic_2019, sc_emitting_regions_2022}. 
We plan to study this in detail in the future exploiting the possibility of fitting both the light curve and the spectra concurrently, using the large set of high-quality data  reported in the 3PC.

\section*{Acknowledgements}

This work has been supported by the grants
PID2021-124581OB-I00, as well as 
the Spanish program Unidad de Excelencia ``María de
Maeztu'' CEX2020-001058-M, 2021SGR00426 of the Generalitat de Catalunya, 
and by MCIU with funding from European Union NextGeneration EU (PRTR-C17.I1).
DIP has been supported by the FPI predoctoral fellowship PRE2021-100290 from the Spanish Ministerio de Ciencia e Innovación and his work has been carried out within the framework of the doctoral program in Physics of the Universitat Autònoma de Barcelona. 
DV is funded by the European Research Council (ERC) under the European Union’s Horizon 2020 research and innovation programme (ERC Starting Grant IMAGINE, No. 948582).

\section*{Data Availability}

No new observational data is herein presented. Any additional theoretical detail required is available from the authors upon reasonable request.

\bibliographystyle{mnras}
\bibliography{sc_lightcurves_spectra_pulsars}

\appendix
\section{Numerical convergence tests}
\label{app:numerical_convergence_tests}

We have performed basic
convergence tests of the peak statistics as a function of the number of injection points along the region, $N_{inj}$. 
While by eye the skymaps look similar, some observers can see different light curves if the region is 
under-resolved (creating artificial noisy features). 
Since we have thousands of light curves, the metric used to evaluate the convergence is the variation of the peak statistics. Table \ref{tab:statistics_different_injection_points} shows the results, for $\Delta R=0.5 R_{lc}$. 
The percentage of observers who see pulsation (detections), and, within them, the 1-peak light curves barely vary. 
However, for $N_{inj} < 16$, there are significant variations among the percentage of 2, 3 and more peaks. For $N_{inj}\geq 21$, such variations stabilize and become less than the estimated systematic $3\%$ error (see text for details). 
Therefore, we have chosen $N_{inj}=21$ injection points, as a compromise between convergence and computational cost.

\begin{table}
    \centering
    \caption{Statistics for different resolution number of injection points, $N_{inj}$ along a region with $\Delta R=0.5~R_{lc}$. As in the main text, we consider about 1530 light curves coming from all the $N_\theta=51$ observers with different values of $\Psi_\Omega$, $\Delta \Psi_\mu$, 10 and 3, respectively, for the pulsar J0205+6449. We indicated the percentage of detected, and, among them, how many have, 1, 2, 3 or more peaks. The percentages can be associated to a statistical error of $\sim 3 \%$ (see main text).}
    \begin{tabular}{c|c|cccc}
        \hline\hline
       $N_{inj}$ & Detected & 1 peak & 2 peaks & 3 peaks & \textgreater 3 peaks \\\hline
        6  & 64.7 & 13.9 & 47.3 & 23.6 & 15.2 \\
        11 & 63.9 & 16.0 & 60.1 & 14.7 & 9.2 \\
        16 & 63.1 & 14.9 & 64.0 & 13.6 & 7.5 \\
        21 & 63.1 & 14.9 & 67.7 & 9.9 & 7.5 \\
        31 & 63.1 & 14.9 & 70.1 & 7.5 & 7.5 \\
        41 & 63.1 & 15.5 & 68.3 & 8.7 & 7.5\\
        \hline
    \end{tabular}
    \label{tab:statistics_different_injection_points}
\end{table}

\section{Details of the peak recognition algorithm}
\label{app:peak_recognition_algorithm}

In order to discard numerical noise (coming from lack of resolution in the trajectories or in the light curve code), we have applied a threshold cut to the intensity of the emission map below which we consider that there is no emission.
Note that this choice also practically mimics the limited signal-to-noise in $\gamma$-ray observations, which make weak signals undetectable (depending of course on the source, its brightness, distance etc.).
We have set this threshold to a $5\%$ of the maximum of the emission map.

In order to calculate the number of peaks seen by each observer and geometry, we normalize each light curve to its maximum intensity. 
Then, any excess of intensity larger than 0.1 (i.e. 10\% of the maximum), is classified as a peak.
In addition we require a minimum phase distance of 0.1 between peaks.
The algorithm then stores the phase position and intensity of each peak of the particular pulsar.
The width of the peaks is computed by fitting the light curve to a (multi-)gaussian distribution and computing the FWHM of each gaussian (the FWHM of a gaussian is $\approx 2.355 \sigma$, where $\sigma$ is the standard deviation of the gaussian).

The choice of the two thresholds and the minimum peak separation allows to have a reliable description of the main features of the synthetic light curves, and we have checked that other reasonable values would not give very different results in the statistics. Note that our algorithm is not directly comparable to the ones used for observational data, since the latter have non-zero background which hides part of the light curves. Indeed, as shown in the main text, the peak statistics changed non negligibly from the 2PC to the 3PC, due to both an enlarged sample and better background treatment.

\label{lastpage}
\end{document}